\documentclass[aos,preprint]{imsart}
\usepackage{mathrsfs}
\usepackage{graphicx}
\usepackage{amsmath}
\usepackage{color}
\usepackage{rotating}
\usepackage{lscape}
\usepackage{graphicx}
\usepackage{bm}
\usepackage{mathrsfs}
\usepackage{booktabs}
\usepackage{caption}
\usepackage{subfig}
\usepackage{slashbox}
\usepackage{float}
\usepackage{amsxtra}
\usepackage{amstext}
\usepackage{amssymb}
\usepackage{latexsym}
\usepackage{dsfont}

\renewcommand{\theequation}{\thesection.\arabic{equation}}

\def\nn{\nonumber}

\def\ha1{\hat \beta_1}

\def\bb0{\delta_\beta}

\def\bsc{\begin{scriptsize}}
\def\esc{\end{scriptsize}}

\def\qed{\qquad$\Box$\medskip}
\input{epsf.sty}
\usepackage{epsfig}

\RequirePackage[OT1]{fontenc}
\RequirePackage{amsthm,amsmath}
\RequirePackage[numbers]{natbib}
\RequirePackage[colorlinks,citecolor=blue,urlcolor=blue]{hyperref}

\startlocaldefs
\numberwithin{equation}{section}
\theoremstyle{plain}

\endlocaldefs
\setattribute{journal}{name}{}

\newtheorem{lemma}{\noindent\mbox{Lemma}}
\newtheorem{theorem}{\noindent\mbox{Theorem}}
\newtheorem{corollary}{\noindent\mbox{Corollary}}
\newtheorem{proposition}{\noindent\mbox{Proposition}}
\renewcommand{\theequation}{\thesection.\arabic{equation}}

\def\be{\begin{equation}}
\def\ee{\end{equation}}
\def\beqr{\begin{eqnarray}}
\def\eeqr{\end{eqnarray}}
\def\beqrs{\begin{eqnarray*}}
\def\eeqrs{\end{eqnarray*}}
\def\bal{\begin{align*}}
\def\eal{\end{align*}}
\def\bet{\begin{theorem}}
\def\eet{\end{theorem}}
\def\bel{\begin{lemma}}
\def\eel{\end{lemma}}
\def\bcor{\begin{corollary}}
\def\ecor{\end{corollary}}
\def\bprop{\begin{proposition}}
\def\eprop{\end{proposition}}

\def\bsc{\begin{scriptsize}}
\def\esc{\end{scriptsize}}

\def\nn{\nonumber}

\newcommand{\E}{\rm E}

\def\bg{\begin{figure}[tbph]\begin{center}}
\def\eg{\end{center}\end{figure}}

\def\bc{\begin{center}}
\def\ec{\end{center}}
\def\bremark{\begin{remark}}
\def\eremark{\end{remark}}

\def\1{\mathbf{1}}

\def\nn{\nonumber}
\def\E{\mbox{E}}

\def\bg{\mathbf{g}}

\begin{document}
\begin{frontmatter}

\title{Test for Temporal Homogeneity of Means in High-dimensional Longitudinal Data}
\runtitle{Tests for Temporal Homogeneity}
\begin{aug}
\author{\fnms{Ping-Shou} \snm{Zhong}\ead[label=e1]{pszhong@stt.msu.edu}}
\and
\author{\fnms{Jun} \snm{Li}\ead[label=e2]{junli@math.kent.edu}}



\affiliation{ Michigan State University and Kent State University}

\address{Department of Statistics and Probability\\
Michigan State University\\
East Lansing, MI 48824,
USA\\
\printead{e1}}

\address{Department of Mathematical Sciences\\
Kent State University\\
Kent, OH 44242,
USA\\
\printead{e2}\\
\phantom{E-mail:\ }}

\end{aug}

\begin{abstract}

This paper considers the problem of testing temporal homogeneity of $p$-dimensional population mean vectors from the repeated measurements of $n$ subjects over $T$ times. 
To cope with the challenges brought by high-dimensional longitudinal data, 
we propose a test statistic that takes into account not only  the ``large $p$, large $T$ and small $n$" situation, but also the complex temporospatial dependence. 
The asymptotic distribution of the proposed test statistic is established under mild conditions. When the null hypothesis of temporal homogeneity is rejected, we further propose a binary segmentation method shown to be consistent 
for multiple change-point identification. Simulation studies and an application to fMRI data are provided to demonstrate the performance of the proposed methods.        

\end{abstract}


\end{frontmatter}

\section{Introduction}

High-dimensional longitudinal data are often observed in modern applications such as genomics studies and neuroimaging studies of brain function. Collected by repeatedly measuring a large number of components from 
a small number of subjects over many time points, the high-dimensional longitudinal data exhibit complex temporospatial dependence: the spatial dependence among the components of each high-dimensional measurement at a particular time point, and the temporal dependence among different high-dimensional measurements collected at different time points. For example, the functional magnetic resonance imaging (fMRI) data are collected by repeatedly measuring the $p$ blood oxygen level-dependent (BOLD) responses from the brains over $T$ times while a small number of subjects are given some task to perform ($p$, $T$ and $n$ are typically at the order of $100,000$, $100$ and $10$, respectively). The fMRI data are characterized by the spatial dependence between the BOLD response in one voxel and a large number of responses measured at neighboring voxels at one time, and the temporal dependence among the BOLD responses of the same subject repeatedly measured at different time points (Ashby, 2011).

This article aims to develop a data-driven and nonparametric method to detect and 
identify temporal changes in a course of high-dimensional time dependent data. Specifically, letting $X_{it}=(X_{it1}, \cdots, X_{tip})^{\prime}$ be a $p$-dimensional random vector observed for the $i$-th subject ($i=1, \cdots, n$) at time $t$ $(t=1, \cdots, T)$, we are interested in testing  
\begin{eqnarray}
&H_0&: \mu_1=\cdots=\mu_T, \qquad \mbox{vs.} \nonumber\\
&H_1&: \mu_1=\cdots=\mu_{\tau_1}\ne \mu_{\tau_1+1}=\cdots=\mu_{\tau_q}\ne \mu_{\tau_q+1}=\cdots=\mu_T, \label{Hypo}
\end{eqnarray}
where $\mu_t$ $(t=1, \cdots, T)$ is a $p$-dimensional population mean vector and $1 \le \tau_1 < \cdots < \tau_q <T$ are $q$ ($q<\infty$) unknown locations of change-points. If the null hypothesis is rejected, we will further estimate the locations of change-points.
The above hypotheses assume that all the individuals come from the same population with
the same mean vectors and change-points. In many applications such as fMRI studies, it is more meaningful to allow the responding mechanism to be different across subjects. This motivates us to further generalize the above hypotheses to (\ref{general-hypo}) where the whole population consists of $G$ ($G>1$) groups, and each group has its own unique means and change-points.  A mixture model is proposed to accommodate such group effect (the details will be introduced in Section 4). 

The classical multivariate analysis of variance (MANOVA) assumes that there exist a finite number ($T<\infty$) of independent normal populations with mean vectors $\mu_1,\cdots,\mu_T$
and common covariance $\Sigma$. In the classical setting with $p<n$, the likelihood ratio test (Wilks, 1932) and Hotelling's $T^2$ test are commonly applied. 
When $p>n$, Dempster (1958, 1960) firstly considered the MANOVA  in the case of two-sample problem. Since then, more methods have been developed in the literature. 
For instance, Bai and Saranadasa (1996) proposed a test by assuming $p/n$ is a finite constant. 
Chen and Qin (2010) further improved the test in Bai and Saranadasa (1996) by proposing a test statistic formulated through the $U$-statistics. 
See also Schott (2007) and Srivastava and Kubokawa (2013). 
Recently, Wang, Peng and Li (2015) proposed a new multivariate test which is able to accommodate heavier tail distributed data. 
Paul and Aue (2014) discussed the applications of random matrix theory in the MANOVA problem. Readers are referred to Fujikoshi et al. (2011) and Hu et al. (2015) for excellent reviews.  

There exist several significant differences between the hypotheses (\ref{Hypo}) considered in this article and the classical MANOVA problem. First, the number of mean vectors $T$
in (\ref{Hypo}) is allowed to diverge to infinity, whereas the typical MANOVA considers the comparison of a finite number of mean vectors. Second, the data considered in this article exhibit complex temporal and spatial 
dependence. 
However, the MANOVA problem typically considers the inference for independent samples without taking into account temporal dependence. Finally, the classical MANOVA problem assumes the homogeneity among subjects but this paper considers the mixture model to accommodate the group effect such that each group is allowed to have its own mean vectors and change-points. Based on the above facts, all of the aforementioned MANOVA methods cannot be applied to the hypotheses (\ref{Hypo}).

In this paper, we propose a new testing procedure for the hypotheses (\ref{Hypo}) under the ``large $p$, large $T$ and small $n$" paradigm. Most importantly, it takes into account both spatial dependence among different components of $X_{it}$, and temporal dependence between $X_{it}$ and $X_{is}$ collected at time points $t \ne s$. The proposed test statistic is constructed in two steps.
In the first step, test statistics are constructed at each $t\in\{1,\cdots, T-1\}$ to distinguish the null from the alternative. In the second step, we choose the maximum of $T-1$ statistics from the first step to make the test free of any tuning parameters and further improve the power. Under some regularity conditions, the maximized statistic is shown to follow the Gumbel distribution if both $T$ and $p$ diverge as $n$ goes to infinite. 
When the null hypothesis of (\ref{Hypo}) is rejected, we further propose a binary segmentation method to identify all the change-points $1 \le \tau_1< \cdots < \tau_q<T$. The proposed method is shown to be consistent for 
the change-point identification by allowing $p$ and $T$ increase as $n$ increases. Moreover, the rate of convergence is established for the proposed change-point estimator, which explicitly includes the effect of dimension 
$p$, time $T$, sample size $n$ as well as the signal-to-noise ratio. 

It is worth mentioning that the current work is different from recent literature on change-point identification under high-dimensionality in several important ways. First, we consider the identification of high-dimensional mean changes that are common to a subgroup of subjects such that inference can be made for a certain population, whereas existing work (e.g., Chen and Zhang, 2015; Jirak, 2015) focuses on change-point identification for high-dimensional time series or panel data  with only one subject ($n=1$). Consequently, the proposed method can establish the consistency of the change-point estimators rather than the ratio consistency (Jirak, 2015). Second, compared with Chen and Zhang (2015) and Jiark (2015), the proposed binary segmentation is computationally efficient. No resampling methods or simulation methods are needed to find the critical values for the change-point identification. Finally, the current work takes into account both temporal and spatial dependence, and the assumptions on dependence structures are very mild. This is different from Chen and Zhang (2015) who assume no temporal dependence, and Jirak (2015) who imposes 
some spatial dependence that requires a natural ordering of $p$ random variables in $X_{it}$. 

The rest of the paper is organized as follows. Section 2 introduces the temporal homogeneity test for the equality of high-dimensional mean vectors at a large number of time points. Its theoretical properties are also investigated. Section 3 
proposes a change-point identification estimator whose rate of convergence is derived. To further identify multiple change-points, we consider a binary segmentation algorithm, which is shown to be consistent. Section 4 extends the 
established temporal homogeneity test and change-point identification method to the mixture model.  Simulation experiment and case study are conducted in Sections 5 and 6 to demonstrate the empirical performance of the proposed 
methods. A brief discussion is given in Section 7. All technical details are relegated to Appendix. Some technical lemmas and additional simulation results are included into a supplementary material.

\section{Temporal Homogeneity Test}
\subsection{Testing Statistic}

We are to propose a test statistic for the hypotheses (\ref{Hypo}). Toward this end, for any $t \in \{1, \cdots, T-1\}$, we first quantify the difference between two sets of mean vectors $\{\mu_{s_1}\}_{s_1=1}^{t}$ and $\{\mu_{s_2}\}_{s_2=t+1}^{T}$ by defining a measure  
\be
M_t=h^{-1}(t)\sum_{s_1=1}^t \sum_{s_2=t+1}^T (\mu_{s_1}-\mu_{s_2})^{\prime} (\mu_{s_1}-\mu_{s_2}), \label{pop-mean}
\ee
where the scale function $h(t)=t(T-t)$. From its definition, $M_t$ is an average of $t(T-t)$ terms, each of which is an Euclidean distance between two population mean vectors chosen before and after a specific 
$t \in \{1, \cdots, T-1\}$. 

Since $M_t=0$ under $H_0$ and $M_t \ne 0$ under $H_1$, it can be used to distinguish the alternative from the null hypothesis. Another advantage of proposing $M_t$ is that it always attains its maximum at one of change-points $\{\tau_1, \cdots, \tau_q\}$ as shown in Lemma 3 in the supplementary material. Thus, it can also be used as a measure for identifying change-points when $H_0$ is rejected (Details will be covered in Section 3). Although there exist other measures for the hypotheses (\ref{Hypo}), some of them might not be designed for identifying change-points. 
For example, Schott (2007)'s test statistic was based on the measure 
$S_{1T}=T\sum_{s=1}^T(\mu_s-\bar{\mu})^{\prime}(\mu_s-\bar{\mu})=\sum_{1\leq s_1<s_2\leq T} (\mu_{s_1}-\mu_{s_2})^{\prime}(\mu_{s_1}-\mu_{s_2})$ where 
$\bar{\mu}=\sum_{s=1}^T\mu_s/T$. It can be shown that $S_{1T}=h(t)M_t+S_{1t}+S_{(t+1)T}$. Note that $S_{1t}$ measures distance among mean vectors before time $t$
and $S_{(t+1)T}$ measures distance among mean vectors after time $t$. Both $S_{1t}$ and $S_{(t+1)T}$ are not informative for the differences 
between the mean vectors $\{\mu_{s_1}\}_{s_1=1}^{t}$ and $\{\mu_{s_2}\}_{s_2=t+1}^{T}$. 

In practice, $M_t$ is unknown. Given a random sample $\{X_{it}=(X_{it1}, \cdots, X_{itp})^{\prime},\\ i=1\cdots, n \, \mbox{and} \,\,  t=1, \cdots, T\}$, it can be estimated by
\begin{eqnarray*}
\hat{M}_t=\frac{1}{h(t)n(n-1)}\sum_{s_1=1}^t \sum_{s_2=t+1}^T \biggl({\sum_{i\ne j}^nX_{is_1}^{\prime}X_{js_1}}+{\sum_{i\ne j}^nX_{is_2}^{\prime}X_{js_2}}-2{\sum_{i\ne j}^nX_{is_1}^{\prime}X_{js_2}} \biggr).\nn
\end{eqnarray*}
Some elementary derivations show that $\mbox{E}(\hat{M}_t)=M_t$. Thus, $\hat{M}_t$ is chosen to be the test statistic for the hypotheses (\ref{Hypo}). 

If $T=2$, the above statistic reduces to the two-sample U-statistics studied by Chen and Qin (2010) for testing the equality of two population means. There are some significant differences between the settings 
considered in current paper and those in Chen and Qin (2010). First, instead of two independent samples in Chen and Qin (2010), we consider high-dimensional time dependent data for testing the equality of more than two 
population mean vectors. There are two types of dependence for consideration: the spatial dependence across the components of $X_{it}$ at a specific time $t$ and the temporal dependence between 
$X_{is}$ and $X_{it}$ with $s \ne t$. Second, although dimension is much larger than sample size in Chen and Qin (2010), $T$ is fixed and equal to 2. Here, we consider the ``large $p$, large $T$ and small $n$" paradigm 
in the sense that both dimension $p$ and time $T$ are much larger than the sample size $n$.

We model $X_{it}$ using a general factor model:
\be
X_{it}=\mu_t+\Gamma_t Z_{i} \qquad \mbox{for} \quad i=1, \cdots, n \quad \mbox{and} \quad t=1, \cdots, T,   \label{model0}
\ee
where $\Gamma_t$ is a $p \times m$ matrix with $m\ge p$ and $\{Z_i\}_{i=1}^n$ are $m$-variate i.i.d. random vectors satisfying $\E (Z_i)=0$, $\mbox{Var} (Z_i)= I_m$, the $m\times m$ identity matrix. If we write $Z_i=(z_{i1}, \cdots, z_{im})^{\prime}$ and let $\Delta$ be a finite constant, we further assume that
\be
\mbox{E}(z_{ik}^4)=3+\Delta, \; \mbox{and} \; \mbox{E}(z_{ik_1}^{l_1}z_{ik_2}^{l_2}\cdots z_{ik_h}^{l_h})=\mbox{E}(z_{ik_1}^{l_1})\mbox{E}(z_{ik_2}^{l_2})\cdots \mbox{E}(z_{ik_h}^{l_h}),\label{factor}
\ee
where $h$ is positive integer such that $\sum_{j=1}^h l_h \le 8$ and $l_1\ne l_2 \ne \cdots \ne l_h$. 

The above models are considered to accommodate the high-dimensional time dependent data. First, (\ref{model0}) enables us to incorporate both spatial and temporal dependence of the data. Let $\delta_{ij}=1$ if $i=j$, and $0$ otherwise. From (\ref{model0}), it immediately follows that 
\be
\mbox{Cov}(X_{is}, X_{jt})=\delta_{ij}\Gamma_s \Gamma_t^{\prime}\equiv \delta_{ij}\Xi_{st}. \nonumber 
\ee
Here $\Xi_{st}$ quantifies the temporal correlation between $X_{is}$ and $X_{it}$ for the same individual measured at different time points $s$ and $t$. Moreover, $\Xi_{st}$ become the covariance matrix $\Sigma_t$ if $s=t$, describing the spatial dependence of $X_{it}$ at time $t$. Second, similar to Chen and Qin (2010) and Bai and Saranadasa (1996), the model (\ref{factor}) allows us to analyze the data beyond commonly assumed Gaussian distribution. 

Define 
\begin{align}
A_{0t}&=\!\sum_{r_1=1}^t\sum_{r_2=t+1}^T (\Gamma_{r_1}-\Gamma_{r_2})^{\prime}(\Gamma_{r_1}-\Gamma_{r_2})\;\mbox{and}\;\nn\\
A_{1t}&=\!\sum_{r_1=1}^t\sum_{r_2=t+1}^T (\mu_{r_1}-\mu_{r_2})^{\prime}(\Gamma_{r_1}-\Gamma_{r_2}). \label{variance}
\end{align}
The following proposition summarizes the variance of the test statistic $\hat{M}_t$.

\bprop 
Under (\ref{model0}), 
\begin{eqnarray}
{\rm Var}(\hat{M}_t)\equiv \sigma_{nt}^2=h^{-2}(t)\Big\{\frac{2}{n(n-1)}{\rm tr}(A_{0t}^2)+\frac{4}{n}||A_{1t}||^2\Big\},\label{mean-var}
\end{eqnarray}
where $A_{0t}$ and $A_{1t}$ are specified in (\ref{variance}), and $\| \cdot \|$ denotes the vector $l^2$-norm. 
\eprop

Specially, $A_{1t}$ becomes a $1\times m$ vector with zeros under $H_0$ of (\ref{Hypo}). Proposition 1 says that the variance of $\hat{M}_t$ under $H_0$ is 
$\sigma_{nt,0}^2=2\mbox{tr}(A_{0t}^2)/\{h^{2}(t)n(n-1)\}$.

\subsection{Asymptotic Distribution of the Proposed Test Statistic }

To establish the asymptotic normality of the proposed test statistic $\hat{M}_t$ at any $t \in \{1, \cdots, T-1 \}$, we require the following condition.

(C1). As $n \to \infty$, $p \to \infty$ and $T \to \infty$, $\mbox{tr}(A_{0t}^4)=o\{\mbox{tr}^2(A_{0t}^2)\}$. In addition, under $H_1$, $A_{1t}A_{0t}^2 A_{1t}^{\prime}=o\{\mbox{tr}(A_{0t}^2)\,\|A_{1t}\|^2\}$. 

Imposing $\mbox{tr}(A_{0t}^4)=o\{\mbox{tr}^2(A_{0t}^2)\}$ is to generalize the condition (3.6) in Chen and Qin (2010) from a fixed $T$ to the diverging $T$ case. Given that $A_{1t}A_{0t}^2 A_{1t}^{\prime}\leq (\max_{k}\lambda_k)\|A_{1t}\|^2$ where $\lambda_k$s are eigenvalues of $A_{0t}^2$, we have $A_{1t}A_{0t}^2 A_{1t}^{\prime}=o\{\mbox{tr}(A_{0t}^2)\,\|A_{1t}\|^2\}$ if $\max_{k}\lambda_k=o\{\mbox{tr}(A_{0t}^2)\}$.  
If the number of non-zero $\lambda_k$s diverges and all the non-zero $\lambda_k$s are bounded, the condition (C1) is easily satisfied.

\bet
\label{th1}
Under (\ref{model0}), (\ref{factor}) and condition (C1), as $n \to \infty$, $p \to \infty$ and $T \to \infty$,
\[
(\hat{M}_t-M_t)/{\sigma_{nt}}  \xrightarrow{d} N(0, 1),
\]
where $\sigma_{nt}$ is defined in (\ref{mean-var}). 
\eet

Specially, under $H_0$, the variance of $\hat{M}_t$ is $\sigma_{nt,0}^2=2\mbox{tr}(A_{0t}^2)/\{h^{2}(t)n(n-1)\}$ with $A_{0t}$ given in (\ref{variance}) and $\hat{M}_t/{\sigma_{nt,0}} \xrightarrow{d} N(0, 1)$. 
In practice, $\sigma_{nt,0}^2$ is unknown. To implement a testing procedure, we estimate $\sigma_{nt,0}^2$ by   
\begin{eqnarray}
\hat{\sigma}_{nt,0}^2=\frac{2}{h^2(t)n(n-1)}\sum_{r_1,s_1=1}^t\sum_{r_2,s_2=t+1}^T \sum_{a,b,c,d \in \{1,2\}}(-1)^{|a-b|+|c-d|}\widehat{\mbox{tr}(\Gamma_{r_b}^{\prime}\Gamma_{r_a}\Gamma_{s_c}^{\prime}\Gamma_{s_d})}, \nonumber
\end{eqnarray} 
where, by defining $P_n^4=n(n-1)(n-2)(n-3)$ to be the permutation number, 
\begin{align}
\widehat{\mbox{tr}(\Gamma_{r_b}^{\prime}\Gamma_{r_a}\Gamma_{s_c}^{\prime}\Gamma_{s_d})}=&
\frac{1}{P_n^4} \sum_{i \ne j \ne k \ne l}^n (X_{i r_a}^{\prime} X_{j r_b} X_{i s_c}^{\prime} X_{j s_d}-X_{i r_a}^{\prime} X_{j r_b} X_{i s_c}^{\prime} X_{k s_d}\nn\\
&\qquad-X_{i r_a}^{\prime} X_{j r_b} X_{k s_c}^{\prime} X_{j s_d}+X_{i r_a}^{\prime} X_{j r_b} X_{k s_c}^{\prime} X_{l s_d}). \label{variance-est} 
\end{align} 
Note that the computational cost of $\hat{\sigma}_{nt,0}^2$ is not an issue. The main reason is
two-fold. First, some simple algebra can be applied to simplify the computation of the summations so that the computation complexity is at the order of $O(n^2T^2p)$. 
Second, the computational cost is mainly due to the size of $n, T$ not $p$, but $n$ and $T$ are 
typically not large in fMRI and genomics applications.

The ratio consistency of $\hat{\sigma}_{nt,0}^2$ is established by the following theorem.  

\bet
Assume the same conditions in Theorem 1. As $n \to \infty$, $p \to \infty$ and $T \to \infty$, 
\[
{\hat{\sigma}_{nt,0}^2}/{\sigma_{nt,0}^2}-1=O_p\big\{ n^{-\frac{1}{2}}{\rm tr}^{-1}(A_{0t}^{2}){\rm tr}^{\frac{1}{2}}(A_{0t}^4)+ n^{-1}\big\}=o_p(1).
\]
\eet

Theorems 1 and 2 lead to a testing procedure that rejects $H_0$ if $\hat{M}_t/ \hat{\sigma}_{nt,0} > z_{\alpha}$ where $z_{\alpha}$ is the upper $\alpha$ quantile of $\mbox{N}(0, 1)$. To implement the testing procedure, we also need to specify $t$, which can be thought as a tuning parameter. Although the type I error of the test will not be affected for any $t \in \{1, \cdots, T-1\}$, the power can be significantly different with respect to different $t$. 
To make our testing procedure free of any tuning parameter, we consider the following test statistic for the hypotheses (\ref{Hypo}):
\be 
\hat{\mathscr{M}}=\max_{0< t/T <1} {\hat{M}_t}/{\hat{\sigma}_{nt,0}}, \label{max-test}
\ee 
which can be readily shown to attain better power than $\hat{M}_t/\hat{\sigma}_{nt,0}$ at any fixed $t \in \{1, \cdots, T\}$ (see the paragraph after Theorem 3 for a proof).

To establish the asymptotic distribution of $\hat{\mathscr{M}}$, we also need (C2) in addition to (C1).

(C2). There exists $\phi(k)$ satisfying $\sum_{k=1}^T\phi^{1/2}(k)<\infty$ such that for any $r, s \in \{1, \cdots, T\}$, $\mbox{tr}(\Xi_{r s}\Xi_{r s}^\prime)\asymp \phi(|r-s|)\mbox{tr}(\Sigma_{r}\Sigma_{s})$. Here $ a \asymp b$ means that 
$a$ and $b$ are of the same order.

The condition (C2) imposes some mild assumption on the temporal dependence among the time series $\{X_{it}\}_{t=1}^T$. 
It basically requires that the time series are weakly dependent to ensure the tightness of the process $\hat{\sigma}_{nt,0}^{-1}\hat{M}_t$ (Billingsley, 1999). 
To establish the weak convergence of $\hat{\mathscr{M}}$, we also define the correlation coefficient 
$r_{nz, uv}=2\mbox{tr}(A_{0u}A_{0v})/\{n(n-1) h(u) h(v) \sigma_{nu,0} \sigma_{nv, 0}\}$ and its limit $r_{z, uv}=\lim_{n \to \infty} r_{nz, uv}.$

\bet  
\label{th3}
Assume (\ref{model0}), (\ref{factor}), (C1), (C2) and $H_0$ of (\ref{Hypo}). As $n\to\infty$ and $p \to \infty$, (i) if $T$ is finite, 
$\hat{\mathscr{M}}  \xrightarrow{d} \max_{0< t/T < 1} Z_t,$
where $Z_t$ is the $t$-th component of $Z=(Z_1, \cdots, Z_{T-1})^{\prime} \sim \mbox{N}(0, R_Z)$ with $R_Z=(r_{z, uv})$;
(ii) if $T \to \infty$ and the maximum eigenvalue of $R_Z$ is bounded, then
$P(\hat{\mathscr{M}}\leq \sqrt{2\log(T)-\log\log(T)+x})\to \exp\big\{-(2\sqrt{\pi})^{-1}\exp(-x/2)\big\}.$
\eet

For the fMRI data analysis, $T$ is typically large and we can apply part (ii) of Theorem \ref{th3}. Specifically,  with $x_{\alpha}=-2\log\{-2\sqrt{\pi}\log (1-\alpha)\}$ defined to be the upper $\alpha$ quantile of the Type I extreme value distribution,   
an $\alpha$-level test rejects $H_0$ of (\ref{Hypo}) if  
$\hat{\mathscr{M}}> {\mathscr{M}}_{\alpha}$ where ${\mathscr{M}}_{\alpha}=\sqrt{2\log(T)-\log\log(T)+x_{\alpha}}$.  Moreover, from Theorems \ref{th1}-\ref{th3}, the lower bound of the power of the test based on $\hat{\mathscr{M}}$ is
\begin{align}
\mbox{P}(\hat{\mathscr{M}} >\mathscr{M}_{\alpha})\ge\max_t\mbox{P}(\frac{\hat{M}_t}{\sigma_{nt,0}}>\mathscr{M}_{\alpha})
 =\max_t\Phi \Big(-\frac{\sigma_{nt, 0}}{\sigma_{nt}}\mathscr{M}_{\alpha}+\frac{M_t}{\sigma_{nt}} \Big),
\end{align}
where $\Phi(\cdot)$ is the cumulative distribution function  of the standard normal. If $\log(T)=o(M_t^2/\sigma_{nt}^2)$ for all $t$, the right hand side of the above expression is the maximum power of the test based on $\hat{M}_t$'s. 
This indicates that the test based on $\hat{\mathscr{M}}$ is more powerful than the test based on the asymptotic normality of $\hat{M}_t$ at a single $t$.

\section{Change-points Identification}

When $H_0$ of (\ref{Hypo}) is rejected, it is very often interesting to further identify the change-points. 
To expedite our analysis, we first consider the simplest case with only one change-point $\tau \in \{1,\cdots, T-1\}$ satisfying the condition $\tau/T=\kappa$ with $0<\kappa<1$. 
It can be shown that $M_t$ attains its maximum at $\tau$, which motivates us to identify the change-point $\tau$ by the following estimator
\be
\hat{\tau}=\arg \max_{0 < t/T < 1} \hat{M}_t. \label{est_CP}
\ee

Let ${v}_{\max}=\max_{1 \le t \le T-1}\max \big\{\sqrt{\mbox{tr}(\Sigma_t^2)}, \sqrt{n (\mu_{1}-\mu_{T})^{\prime}\Sigma_t(\mu_{1}-\mu_{T})} \big\}$
and 
$\delta^2=(\mu_{1}-\mu_{T})^{\prime}(\mu_{1}-\mu_{T})$. 
The following theorem establishes the rate of convergence for the change point estimator $\hat{\tau}$. 

\bet
Assume that a change-point $\tau \in \{1,\cdots, T-1\}$ satisfies $\tau/T=\kappa$ with $0<\kappa<1$,
$(\mu_{1}-\mu_{T})^{\prime}\Xi_{r s}(\mu_{1}-\mu_{T})\asymp \phi(|r-s|)(\mu_{1}-\mu_{T})^{\prime}\Sigma_{r}(\mu_{1}-\mu_{T})$,
where $\phi(\cdot)$ is defined in condition (C2). 
Under (\ref{model0}), (\ref{factor}), (C1) and (C2), as $n \to \infty$, 
\[
\hat{\tau}-\tau=O_p\big\{\sqrt{T\log(T)}\,\,{v}_{\max}/(n \,\delta^2) \big\}.
\]
\eet

Theorem 4 shows that $\hat{\tau}$ is consistent to $\tau$ if $n \delta^2 /\{{v}_{\max} \sqrt{T\log(T)}\} \to \infty$, where $n\delta^2$ is a measure of signal and ${v}_{\max}$ is associated with noise. Most importantly, it explicitly demonstrates the contributions of dimension $p$, time $T$ and sample size $n$ to the rate of convergence. First, if both $p$ and $T$ are fixed, $\hat{\tau}-\tau=O_p(n^{-1/2})$ as $n \to \infty$. Second, if $p$ is fixed but $T$ diverges as $n$ increases, $\hat{\tau}-\tau=O_p(\sqrt{T\log(T)/n})$.  
Last but not least, if both $p$ and $T$ diverge as $n$ increases, 
the convergence rate can be faster than $O_p(\sqrt{T\log(T)/n})$. To appreciate this, we consider a special setting where $X_{it}$ in (\ref{model0}) has the identity covariance $\Sigma_t=I_p$, 
the non-zero components of $\delta^2$ are equal and fixed, and the number of non-zero components is $p^{1-\beta}$ for $\beta \in (0,1)$. Under such setting, 
\[
\hat{\tau}-\tau=O_p\biggl (\frac{\{T\log(T)\}^{1/2}}{\min\{np^{1/2-\beta}, n^{{1}/{2}} \,p^{(1-\beta)/2}\}} \biggr),
\]
which is faster than the rate $O_p\{\sqrt{T\log(T)/n}\}$ if  $n^{1/2}p^{1/2-\beta}\to\infty$.

Next, we consider that there exist more than one change-point. To identify these change-points, we first define some notation. Let $\mathbb{S}= \{1 \le \tau_1 < \cdots < \tau_q <T\}$ be a set containing all $q$ ($q \ge 1$) change-points. For any $t_1, t_2 \in \{1, \cdots, T\}$ satisfying $t_1 < t_2$, let $\hat{\mathscr{M}}[t_1, t_2]$ and $\mathscr{M}_{\alpha_n}[t_1, t_2]$ denote the maximum test statistic in (\ref{max-test}) and the corresponding upper $\alpha_n$ quantile, calculated based on data collected between the time points $t_1$ and $t_2$. Lemma 3 in  supplementary material shows that $M_t$ in (\ref{pop-mean}) always attains its maximum at one of the change-points, which motivates us to identify all change-points by the following binary segmentation algorithm (Venkatraman, 1992).
\begin{enumerate}
\item[(1).] Check if $\hat{\mathscr{M}}[1, T] \le \mathscr{M}_{\alpha_n}[1, T]$. If yes, then no change-point is identified and stop. Otherwise, a change-point $\hat{\tau}_{(1)}$ is selected by 
$\hat{\tau}_{(1)}=\mbox{arg} \max_{1\le t \le T-1} \hat{M}_t$,  
and included into $\hat{\mathbb{S}}=\{ \hat{\tau}_{(1)} \}$;
\item[(2).] Treat $\{1, \hat{\tau}_{(1)}, T\}$ as new ending points and first check if $\hat{\mathscr{M}}[1, \hat{\tau}_{(1)}] \le \mathscr{M}_{\alpha_n}[1, \hat{\tau}_{(1)}]$. If yes, no change-point is selected from time 1 to $\hat{\tau}_{(1)}$. Otherwise, one change-point is selected by
$\hat{\tau}^1_{(2)}=\mbox{arg} \max_{1\le t \le \hat{\tau}_{(1)}-1} \hat{M}_t$, 
and update $\hat{\mathbb{S}}$ by adding $\hat{\tau}^1_{(2)}$. Next check if $\hat{\mathscr{M}}[\hat{\tau}_{(1)}+1, T] \le \mathscr{M}_{\alpha_n}[\hat{\tau}_{(1)}+1, T]$. If yes, no time point is selected from time $\hat{\tau}_{(1)}+1$ to $T$. Otherwise, one change-point is selected by $\hat{\tau}^2_{(2)}=\mbox{arg} \max_{\hat{\tau}_{(1)}+1 \le t \le T-1} \hat{M}_t$, 
and $\hat{\mathbb{S}}$ is updated by including $\hat{\tau}^2_{(2)}$. If no any change-point has been identified from both $[1, \hat{\tau}_{(1)}]$ and $[\hat{\tau}_{(1)}+1, T]$, then stop. Otherwise, rearrange $\hat{\mathbb{S}}$ by sorting its elements from smallest to largest and update ending points by $\{1, \hat{\mathbb{S}}, T\}$;
\item[(3).]  Repeat step 2 until no more change-point is identified from each time segment, and obtain the final set $\hat{\mathbb{S}}$ as an estimate of the set $\mathbb{S}$.
\end{enumerate}

Define $\tau_0=1$ and $\tau_{q+1}=T$. Let $I_t$ be any time interval of the form $I_t=[\tau_i+1, \tau_j]$ with $i+1<j$ that contains at least one change-point $\tau_i$ for $i \in \{1, \cdots, q\}$, and define the smallest maximum signal-to-noise ratio among all time intervals $I_t$ to be $\mathscr{R}^*=\min_{I_t} \max_{\tau_i \in I_t} M[I_t]/\sigma_n[I_t]$ where $M[I_t]$ and $\sigma_n[I_t]$ are (\ref{pop-mean}) and (\ref{mean-var}) specified in $I_t$, respectively. To establish the consistency of $\hat{\mathbb{S}}$ obtained from the above binary segmentation algorithm, we need the following condition in addition to (C1) and (C2).

(C3). As $T \to \infty$, $\tau_i/T$ converges to $\kappa_i$ for $i=1, \cdots, q$ with fixed $q \ge 1$, satisfying $0 < \kappa_1< \cdots < \kappa_q <1$.

\bet
Assume (\ref{model0}), (\ref{factor}), (C1)-(C3), and $\mathscr{R}^*$ diverges such that the upper $\alpha_n$-quantile of the Gumbel distribution 
$\mathscr{M}_{\alpha_n}=o(\mathscr{R}^*)$ as $\alpha_n \to 0$. Furthermore, $v_{\max}[I_t]=o\{n\delta^2[I_t]/\sqrt{T\log(T)}\}$ for all $I_t$ that contains at least one change-point. Then,   
$\hat{\mathbb{S}} \xrightarrow{p} \mathbb{S}, $ as $n \to \infty$ and $T \to \infty$. 
\eet

%
%
%
%

\section{An Extension to Mixture Models}

Thus far we focus on temporal homogeneity detection by assuming that all subjects in the sample come from a population with the same change-points. In fMRI experiments, if different subjects choose different strategies 
to solve the same task, the patterns activated by stimuli will be different across subjects (Ashby, 2011). Analytically, it is more attractive to consider that subjects show the same activation pattern within each group, but 
different patterns across groups. 

In this section, we will generalize the approaches developed in the last two sections to accommodate such group effect. Instead of the model (\ref{model0}) considered in Section 2, we assume that the data follow a mixture model 
\be
X_{it}=\sum_{g=1}^G \Lambda_{ig}\mu_{gt}+\Gamma_t Z_i,
\label{mixture-of-means}
\ee
where independent of $\{Z_i \}_{i=1}^n$, $(\Lambda_{i1},\cdots, \Lambda_{iG})$ follows a multinomial distribution with parameters 1 and $p=(p_1,\cdots,p_G)$.
This suggests that $\sum_{g=1}^G\Lambda_{ig}=1$ with $\Lambda_{ig}\in\{0,1\}$, and
$\mbox{P}(\Lambda_{ig}=1)=p_g$ satisfying $\sum_{g=1}^G p_g=1$ with the number of groups $G \ge 1$. 
Note that the above model implies that $i$-th subject only belongs to one of $G$ groups. 
The mixture model is more general because (\ref{model0}) is a special case of (\ref{mixture-of-means}) if there is only one group ($G=1$).  

The mixture morel (\ref{mixture-of-means}) is also flexible because it allows each group to have its own population mean vectors $\{\mu_{gt}\}_{t=1}^T$ for $g=1, \cdots, G$. In analogy to (\ref{Hypo}), we want to know whether there exist some change-points within some groups by testing  
\begin{align}
H_0^*: \mu_{g1}&=\mu_{g2}=\cdots=\mu_{gT}\;\;\mbox{for all $1\leq g\leq G$ \quad vs.}\nn \\
H_1^*: \mu_{g1}&=\cdots=\mu_{g\tau_1^{(g)}}\neq \mu_{g(\tau_1^{(g)}+1)}=\cdots=\mu_{g\tau_{q_g}^{(g)}}\neq \mu_{g(\tau_{q_g}^{(g)}+1)}=\cdots=\mu_{gT}\nn\\
    &\;\mbox{for some $g$}.\label{general-hypo} 
\end{align}
If $H_0^*$ is rejected, we further identify $\{\tau_1^{(g)},\tau_2^{(g)}\cdots,\tau_{q_g}^{(g)}\}_{g=1}^G$, the collection of $q$ ($q=\sum_{g=1}^G q_g$) change-points from $G$ groups. 

Toward this end, we first evaluate the mean and variance of the test statistic $\hat{M}_t$ under the mixture model (\ref{mixture-of-means}). Similar to Proposition 1, the mean is $E(\hat{M}_t)=\tilde{M}(t)=
h^{-1}(t)\sum_{r_1=1}^t\sum_{r_2=t+1}^{T}(\tilde{\mu}_{r_1}-\tilde{\mu}_{r_2})^{\prime}(\tilde{\mu}_{r_1}-\tilde{\mu}_{r_2})$ with $\tilde{\mu}_{r_i}=\sum_{g=1}^G p_g\mu_{g r_i}$ for $i=1, 2$. The variance of $\hat{M}_t$ is
\be
\mbox{Var}(\hat{M}_t)\equiv \tilde{\sigma}_{nt}^2= \frac{2}{n(n-1)h^{2}(t)}\{\mbox{tr}(A_{0t}^2)+\tilde{A}_{3t}\}+\frac{4}{nh^{2}(t)}\{||\tilde{A}_{1t}||^2+\tilde{A}_{2t}\}, \label{mix_var}
\ee
where $A_{0t}$ is defined in (\ref{variance}), $\tilde{A}_{1t}=\sum_{r_1=1}^t \sum_{r_2=t+1}^T(\tilde{\mu}_{r_1}-\tilde{\mu}_{r_2})^{\prime}(\Gamma_{r_1}-\Gamma_{r_2})$. In addition, with $\delta_{g_1 g_2 r_i}=\mu_{g_1 r_i}-\mu_{g_2 r_i}$ for $i=1, 2$, 
\begin{align}
\tilde{A}_{2t}&=\sum_{g_1<g_2}^G  p_{g_1} p_{g_2} \Big\{\sum_{r_1=1}^t \sum_{r_2=t+1}^T(\delta_{g_1 g_2 r_1}-\delta_{g_1 g_2 r_2})^{\prime}(\tilde{\mu}_{r_1}-\tilde{\mu}_{r_2}) \Big\}^2\;\;\mbox{and}\;\;\nn\\
\tilde{A}_{3t}&=\sum_{g_1<g_2, g_3<g_4}^G p_{g_1} p_{g_2}p_{g_3} p_{g_4} \Big\{\sum_{r_1=1}^t \sum_{r_2=t+1}^T(\delta_{g_1 g_2 r_1}-\delta_{g_1 g_2 r_2})^{\prime}(\delta_{g_3 g_4 r_1}-\delta_{g_3 g_4 r_2})
\Big\}^2.\nn
\end{align}

It is worth discussing some special cases of (\ref{mix_var}). First, if there is only one group ($G=1$), it can be shown that $\tilde{A}_{2t}=\tilde{A}_{3t}=0$, and $\tilde{A}_{1t}=A_{1t}$ defined in (\ref{variance}). Therefore, the variance formulated in Proposition 1 is a special case of the variance (\ref{mix_var}) under the mixture model. Second, under $H_0^*$ of (\ref{general-hypo}),  
$\tilde{\sigma}_{nt,0}^2\equiv\mbox{Var}(\hat{M}_t)=2\mbox{tr}(A_{0t}^2)/\{n(n-1)h^{2}(t)\}$ because $\tilde{A}_{1t}=\tilde{A}_{2t}=\tilde{A}_{3t}=0$. The unknown $\tilde{\sigma}_{nt,0}^2$ can be estimated by 
$$
\widehat{\tilde{\sigma}}_{nt,0}^2=\frac{2}{h^2(t)n^2(n-1)^2}\sum_{i \ne j}^n\Big\{\sum_{r_1=1}^t\sum_{r_2=t+1}^T \sum_{a,b\in \{1,2\}}(-1)^{|a-b|}
 X_{i r_a}^{\prime} X_{j r_b}\Big\}^2.
$$

Similar to $\hat{\mathscr{M}}$ given by (\ref{max-test}), we define 
$\tilde{\mathscr{M}}=\max_{1\le t \le T-1} \hat{M}_t/\widehat{\tilde{\sigma}}_{nt,0}$. The temporal homogeneity detection and 
identification procedures developed in Sections 2 and 3 can be extended to testing the hypothesis in (\ref{general-hypo}) by replacing $\hat{\mathscr{M}}$ with $\tilde{\mathscr{M}}$.
Furthermore, the asymptotic results in Theorem 1-5 can be established for the mixture model (\ref{mixture-of-means}) under some regularity 
conditions. Due to the space limitation, we only demonstrate the empirical performance under the mixture model through simulation studies and leave explorations of the theoretical results to future study.

\section{Simulation Studies}

In this section, we will evaluate the finite sample performance of the methods proposed in Sections 2--4. 

\subsection{Test for the Homogeneity of Means }

We first evaluate the numerical performance of the test procedure proposed in Section 2.
The random sample $\{X_{it}\}$ for $i=1, \cdots, n$ and $t=1, \cdots, T$, were generated from the following
multivariate linear process
\be
X_{it}=\mu_t+\sum_{l=0}^J Q_{lt}\,\epsilon_{i(t-l)}, \label{data_g}
\ee 
where $\mu_t$ is the $p$-dimensional population mean vector at time $t$, $Q_{lt}$ is a $p \times p$ matrix and $\epsilon_{it}$ is $p$-variate normally distributed with mean $0$ and identity covariance $\mbox{I}_p$. The model was considered to account for both time dependence of $X_{it}$ and $X_{is}$ at $t\ne s$, and spatial dependence among the $p$-components of $X_{it}$ at a specific time $t$. Specifically, it can be seen that 
$\mbox{Cov}(X_{it}, X_{is})= \sum_{l=t-s}^J Q_{lt} Q_{(l-t+s)s}$ if $t-s\le J$ and $\mbox{Cov}(X_{it}, X_{is})=0$ otherwise.
Note that $J$ is used to control the level of dependence. As $J$ increases, the temporal dependence among $\{X_{it}\}_{t=1}^T$ becomes stronger.

In the simulation, we chose $J=2$ and $Q_{lt}=\{0.5^{|i-j|}\mbox{I}(|i-j|<p/2)/(J-l+1)\}$ for $i, j=1, \cdots, p$, and $l=0, 1, 2$. 
To evaluate the empirical size of the proposed test, we simply chose $\mu_t=0$ for all $t$ under $H_0$ of (\ref{Hypo}). 
Under $H_1$, we considered one change-point located at $0.4 \cdot T$ such as $\mu_t=0$ for $t=1, \cdots, 0.4T$ and $\mu_t=\mu$ for $t=0.4T+1, \cdots, T$. The non-zero mean vector $\mu$ had $[p^{0.7}]$ non-zero components which were uniformly and randomly drawn from $p$ coordinates $\{1, \cdots, p\}$. Here, $[a]$ denotes the integer part of $a$. The magnitude of non-zero entry of $\mu$ was controlled by a constant $\delta$ multiplied by a random sign. The effect of sample size, dimensionality, and length of time series on the performance of the proposed testing procedure was demonstrated by different combinations of $n \in \{30, 60, 90\}$, $p \in \{50, 100, 200\}$ and $T \in \{50, 100, 150\}$. The nominal significance level was chosen to be $0.05$. All the simulation results were obtained based on 1000 replications. 

Table \ref{case1} summarizes the empirical performance of the proposed procedure for testing the homogeneity of means. 
All the empirical sizes ($\delta=0$) were well controlled under the nominal significance level $0.05$ although some of them were relatively conservative. 
This is largely due to the slow convergence of the Gumbel distribution. Furthermore, the empirical powers increased as $p$, $T$ and $n$ increased, 
which confirms the theoretical findings of the proposed testing procedure.

\begin{table}[t]
\tabcolsep 4pt
\begin{center}
\caption{Empirical sizes and powers of the proposed test for homogeneity of means under different combinations of $n$, $p$ and $T$.} 
\label{case1}
\begin{tabular}{ccccccccccccccc}
\hline  &\multicolumn{2}{c}{}&\multicolumn{4}{c}{$T=50$} &\multicolumn{4}{c}{$T=100$} &\multicolumn{4}{c}{$T=150$}\\[1mm]
\cline{5-7} \cline{9-11} \cline{13-15}$\delta$& &n& &$p=50$ & $100$ & $200$ & &$p=50$ & $100$ & $200$ &  & $p=50$ 
& $100$& $200$ \\\hline
    & &$30$ & & $0.040$ & $0.033$ & $0.028$ & &$0.049$& $0.030$ & $0.017$& &$0.034$& $0.042$& $0.024$ \\
 0   & &$60$ & & $0.052$ & $0.036$ & $0.021$ & &$0.033$& $0.031$ & $0.021$& &$0.029$& $0.014$& $0.015$ \\
    & &$90$ & & $0.050$ & $0.032$ & $0.022$ & &$0.033$& $0.024$ & $0.017$& &$0.051$& $0.031$& $0.018$ \\\hline
    & &$30$ & & $0.117$ & $0.121$ & $0.122$ & &$0.141$& $0.157$ & $0.216$& &$0.186$& $0.209$& $0.276$ \\
 0.2   & &$60$ & & $0.309$ & $0.362$ & $0.504$ & &$0.523$& $0.738$ & $0.833$& &$0.731$& $0.884$& $0.982$ \\
    & &$90$ & & $0.578$ & $0.790$ & $0.922$ & &$0.918$& $0.995$ & $0.999$& &$0.993$& $1.000$& $1.000$ \\        
\hline
    & &$30$ & & $0.378$ & $0.474$ & $0.633$ & &$0.648$& $0.826$ & $0.938$& &$0.860$& $0.954$& $0.992$ \\
 0.3   & &$60$ & & $0.956$ & $0.994$ & $1.000$ & &$1.000$& $1.000$ & $1.000$& &$1.000$& $1.000$& $1.000$ \\
    & &$90$ & &  1.000 & 1.000 & 1.000 & & 1.000 & 1.000 & 1.000 & & 1.000 & 1.000 & 1.000 \\\hline 
\end{tabular}
\end{center}
\end{table}

%

\subsection{Change-Point Identification }

Simulation experiments were also conducted to evaluate the change-point identification procedure proposed in Section 3. We generated data using similar setup for change-point testing in the last subsection, but we considered two 
change-points at $0.4 \cdot T$ and $0.7 \cdot T$ such as $\mu_t=0$ for $t=1, \cdots, 0.4T$, $\mu_t=\mu$ for $t=0.4T+1, \cdots, 0.7T$ and $\mu_t=0$ for $t=0.7T+1, \cdots, T$. 
Again, the non-zero mean vector $\mu$ had $[p^{0.7}]$ non-zero components which were uniformly and randomly drawn from $\{1, \cdots, p\}$. 
The non-zero entry of $\mu$ was $\delta=0.5$ and $\delta=0.6$, respectively, multiplied by a random sign.    

\begin{figure}[t!]
\begin{center}
\includegraphics[width=0.4\textwidth,height=0.4\textwidth]{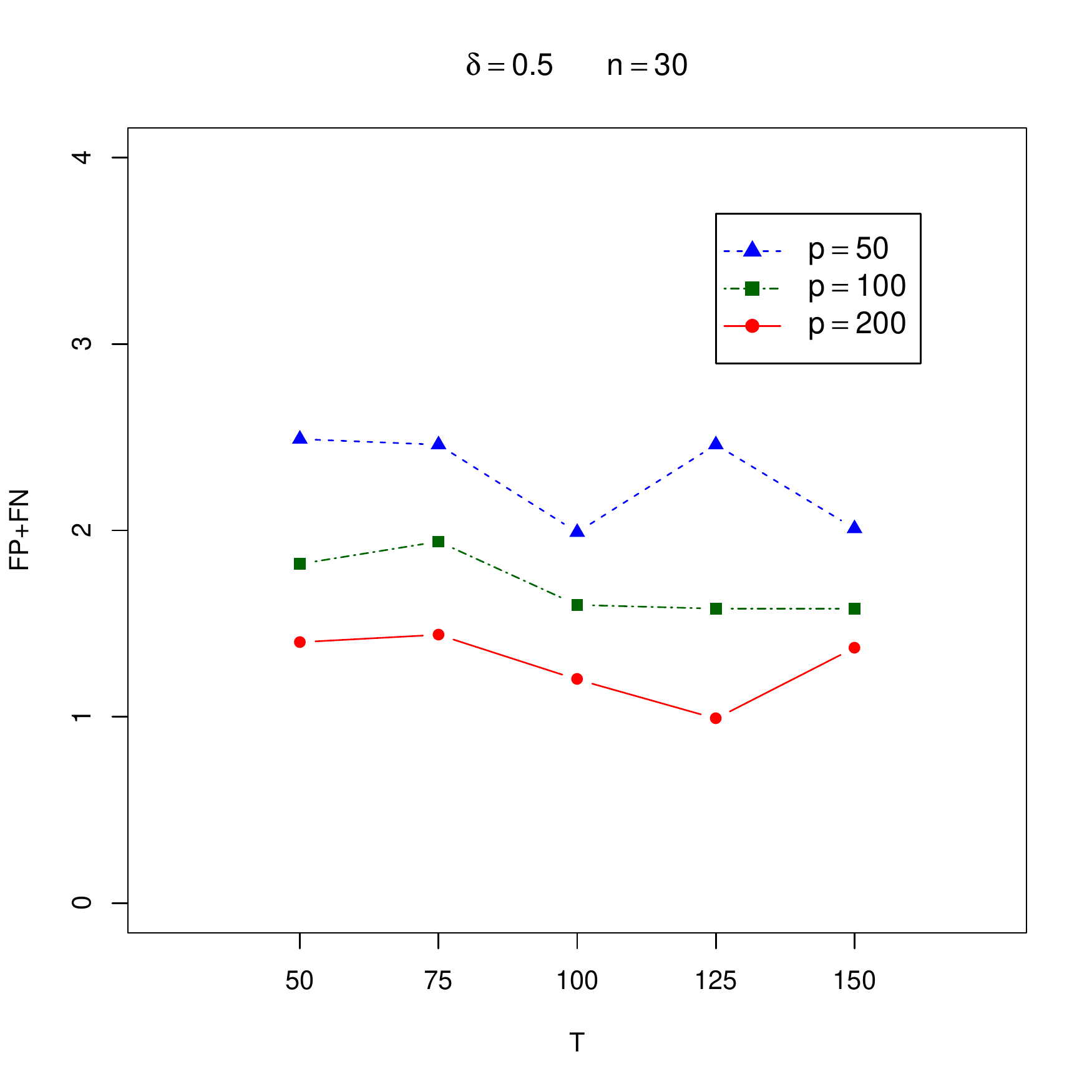}
\includegraphics[width=0.4\textwidth,height=0.4\textwidth]{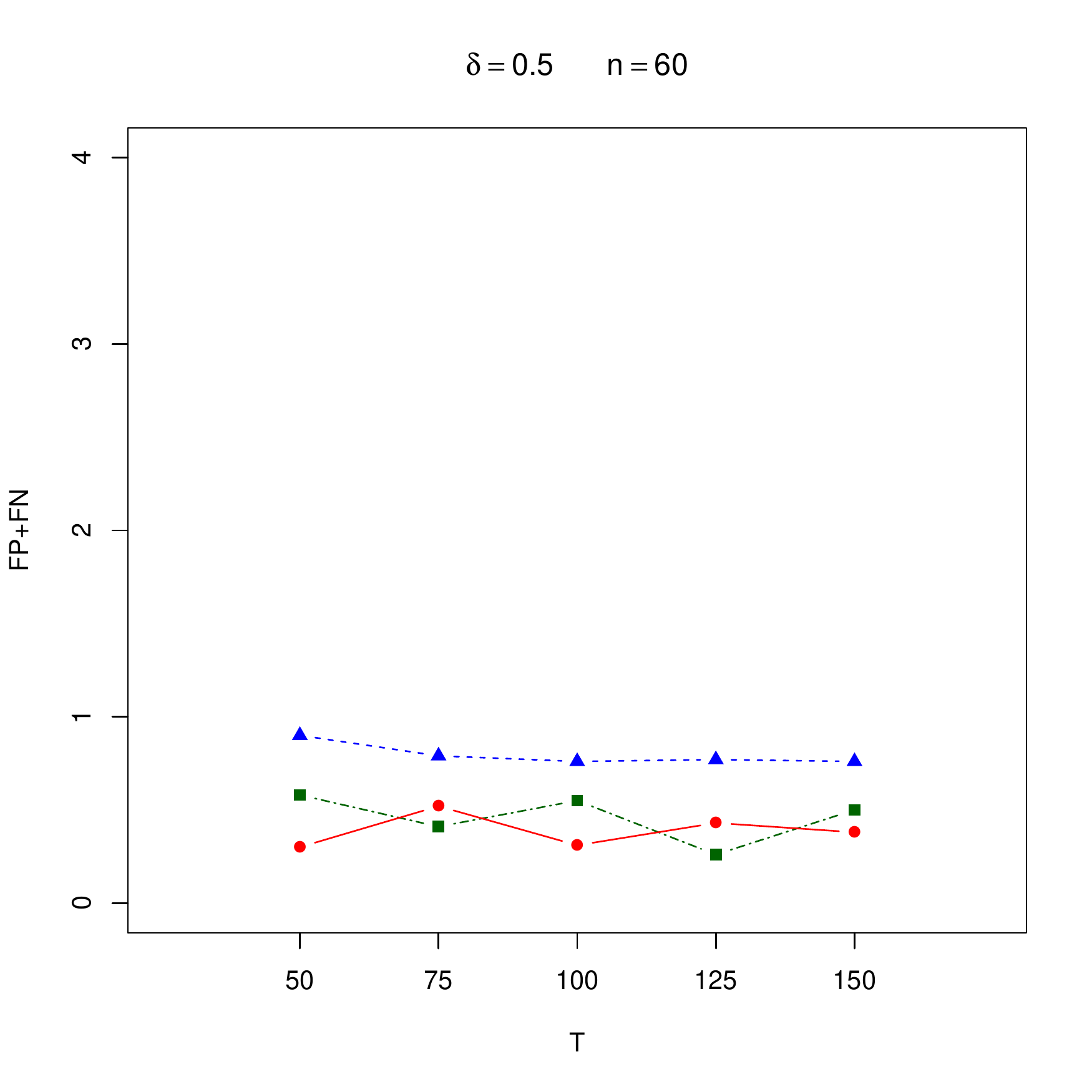}\\
\includegraphics[width=0.4\textwidth,height=0.4\textwidth]{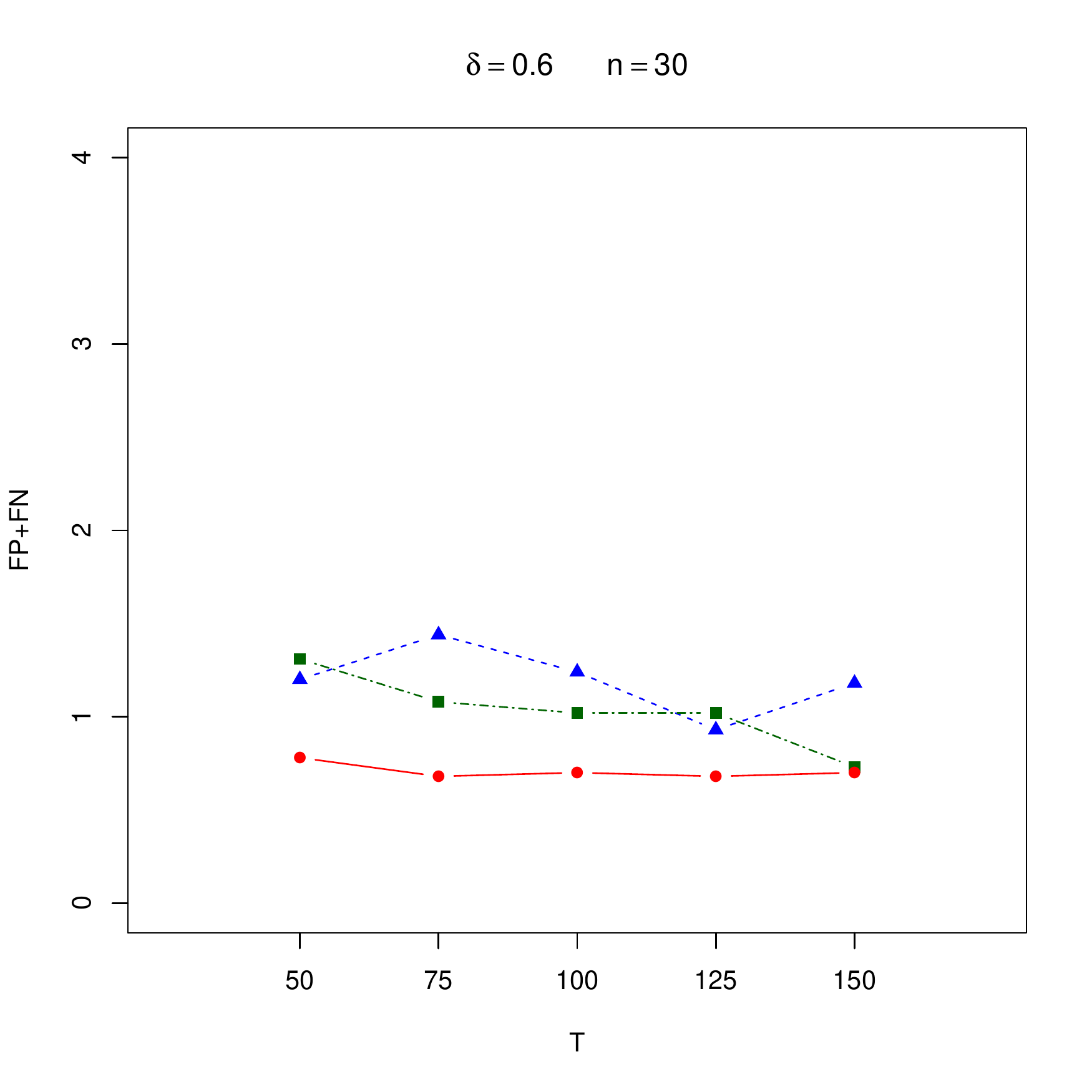}
\includegraphics[width=0.4\textwidth,height=0.4\textwidth]{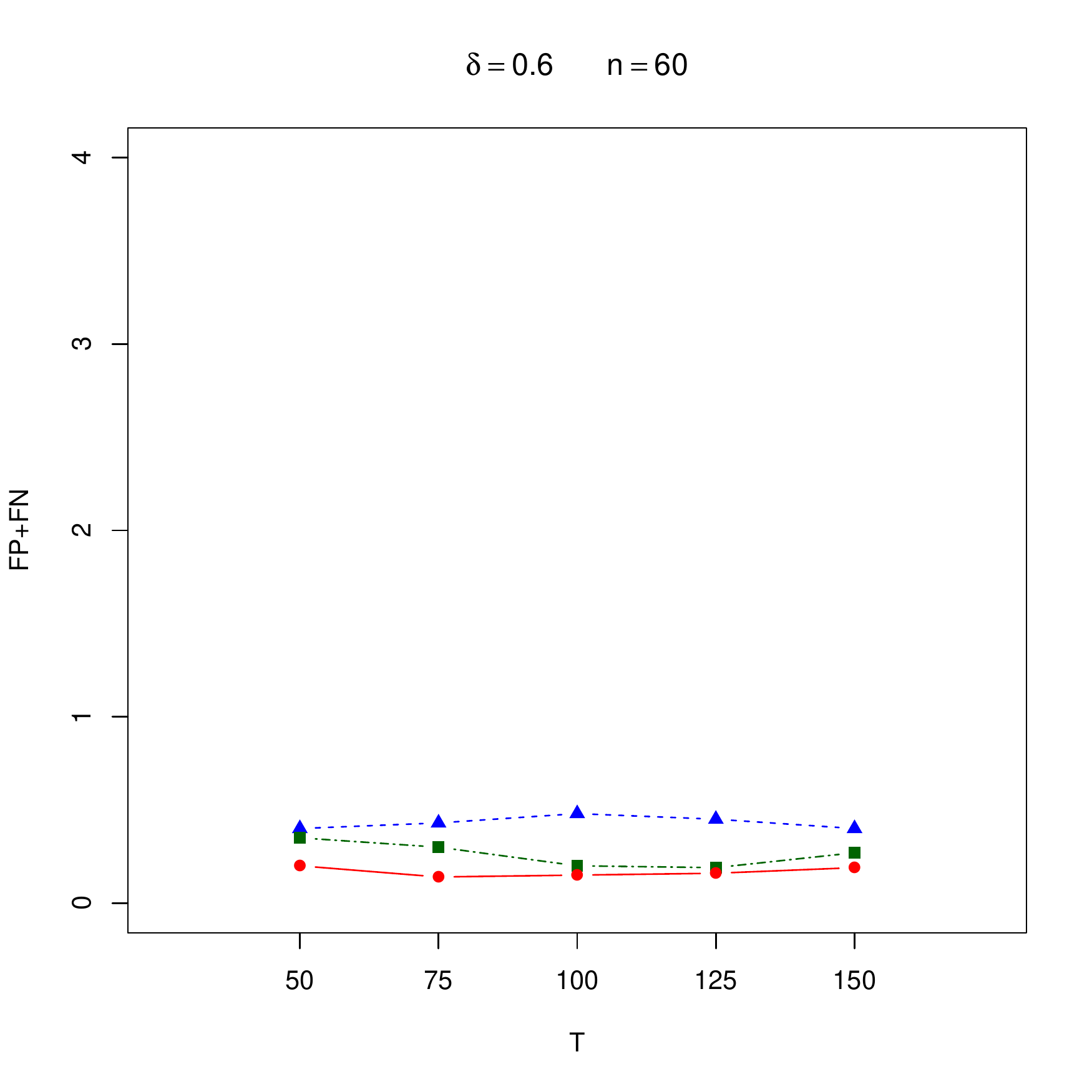}
\caption{The average FP+FN under different combinations of signal strength $\delta$, dimension $p$, time $T$ and sample size $n$. The total number of change-points are set to be $2$. }
\label{fig2}
\end{center}
\end{figure}

There are two types of errors for change-point identification: the false positive (FP) and the false negative (FN).
The FP means that a time point without changing the mean is wrongly identified as a change-point, and the FN refers that a change-point is wrongly treated as a time point without changing the mean. 
The accuracy of the proposed change-point identification was measured by the sum of FP and FN. Simulation results were obtained based on 100 replications.

Figure \ref{fig2} demonstrates the FP+FN associated with the proposed change-point identification procedure under different combinations of $\delta$, $p$, $T$ and $n$. 
More specifically, the average FP+FN decreased as $\delta$ increased with fixed $p$, $T$ and $n$. 
Also the FP+FN decreased as either $p$ increased with fixed $\delta$, $T$ and $n$, or $n$ increased with fixed $\delta$, $p$ and $T$. 
In the supplementary material, we also summarize the performance using the number of true positives (TP). The results show that the TP identified by the proposed procedure converged to the number of change-points (see supplementary material for details).

We also conducted simulation studies for the proposed change-point detection and identification methods with non-Gaussian data. Instead of using the normally distributed $\epsilon_{it}$ in (\ref{data_g}), 
we considered the centralized Gamma(4, 0.5). The results were similar to those given in Table 1 and Figure 1, which shows that the proposed test is presumably nonparametric in the sense that it does not rely on the Gaussian data. Due to the space limitation, the results are reported in the supplementary material.

\subsection{Detection and Identification Under the Mixture Model}

To evaluate the performance of the proposed methods under the mixture model (\ref{mixture-of-means}), we generated the data from the following model with three groups:
\be
X_{it}=\sum_{g=1}^3 \Lambda_{ig}\mu_{gt}+\sum_{l=0}^J Q_{lt}\,\epsilon_{i(t-l)}, 
\label{mixture-of-means-simu}
\ee
where $(\Lambda_{i1},\Lambda_{i2}, \Lambda_{i3})$ follows a multinomial distribution with parameters 1 and $p=(p_1, p_2,p_3)$.
satisfying $P(\Lambda_{ig}=1)=p_g$ for $g=1,2$ and 3. In the simulation, we set $(p_1,p_2,p_3)=(0.3,0.3,0.4)$. Among three groups, we considered two change-points $\tau_1=0.4 \cdot T$ and $\tau_2=0.7 \cdot T$. 
Specifically, for the first group ($g=1$), $\mu_{1t}=0$ for $1 \leq t\leq \tau_1$ and $\mu_{1t}=\mu_1$ for $\tau_1+1\leq  t\leq T$, where $\mu_{1}$ had $[p^{0.7}]$ non-zero components drawn uniformly and 
randomly from $\{1, \cdots, p\}$. The magnitude of non-zero entry of $\mu_{1}$ was $\delta_1$ multiplied by a random sign. 
For the second group  ($g=2$), the mean vectors $\mu_{2t}$ were obtained similarly to those for the first group except that we changed $\tau_1$ to $\tau_2$, and $\delta_1$ to $\delta_2$.
For the third group ($g=3$), we set $\mu_{3t}=0$ for $1 \leq t\leq \tau_1$, $\mu_{3t}$ equal to the non-zero mean vectors similar to those in group 2 for $\tau_1+1 \le t \le \tau_2$, 
and $\mu_{3t}=\mu_3$ for $\tau_2+1 \leq t\leq T$ where $\mu_3$ were generated similarly to that in the first group except that we changed $\delta_1$ to $\delta_3$.

\begin{table}[t]
\tabcolsep 3pt
\begin{center}
\caption{Empirical powers of the proposed test under the mixture model with different combinations of $n$, $p$ and $T$.} \label{case2}
\begin{tabular}{ccccccccccccccc}
\hline  &\multicolumn{2}{c}{}&\multicolumn{4}{c}{$T=50$} &\multicolumn{4}{c}{$T=100$} &\multicolumn{4}{c}{$T=150$}\\[1mm]
\cline{5-7} \cline{9-11} \cline{13-15}
$(\delta_1,\delta_2,\delta_3)$& &n& &$p=50$ & $100$ & $200$ & &$p=50$ & $100$ & $200$ &  & $p=50$ 
& $100$& $200$ \\\hline
          & &$30$ & &  0.094 & 0.088 & 0.123 & & 0.119 & 0.142 & 0.179 & & 0.139 & 0.174 & 0.241\\
 (0.25, 0.35, 0.4)  & &$60$ & &  0.300 & 0.349 & 0.445 & & 0.463 & 0.592 & 0.712 & & 0.618 & 0.752 & 0.882\\
          & &$90$ & &  0.533 & 0.690 & 0.814 & & 0.817 & 0.928 & 0.979 & & 0.929 & 0.981 & 0.998\\\hline
          & &$30$ & &  0.691 & 0.816 & 0.907 & & 0.892 & 0.957 & 0.987 & & 0.946 & 0.991 & 0.999\\
(0.5, 0.7, 0.8)   & &$60$ & &  0.997 & 1.000 & 1.000 & & 1.000 & 1.000 & 1.000 & & 1.000 & 1.000 & 1.000\\
          & &$90$ & &  1.000 & 1.000 & 1.000 & & 1.000 & 1.000 & 1.000 & & 1.000 & 1.000 & 1.000\\ \hline
\end{tabular}
\end{center}
\end{table}

We first evaluate the proposed test under the mixture model (\ref{mixture-of-means-simu}). Since the empirical sizes under the mixture model were very similar to those in Table \ref{case1}, 
we only report the empirical powers in Table \ref{case2}. The patterns are very similar to what we observed in Table \ref{case1}. 
We also observe that the empirical powers increased as $(\delta_1,\delta_2, \delta_3$), or $n$, $p$ and $T$ increased. This suggests that the proposed test procedure is consistent under the mixture model.

\begin{figure}[t!]
\begin{center}
\includegraphics[width=0.4\textwidth,height=0.4\textwidth]{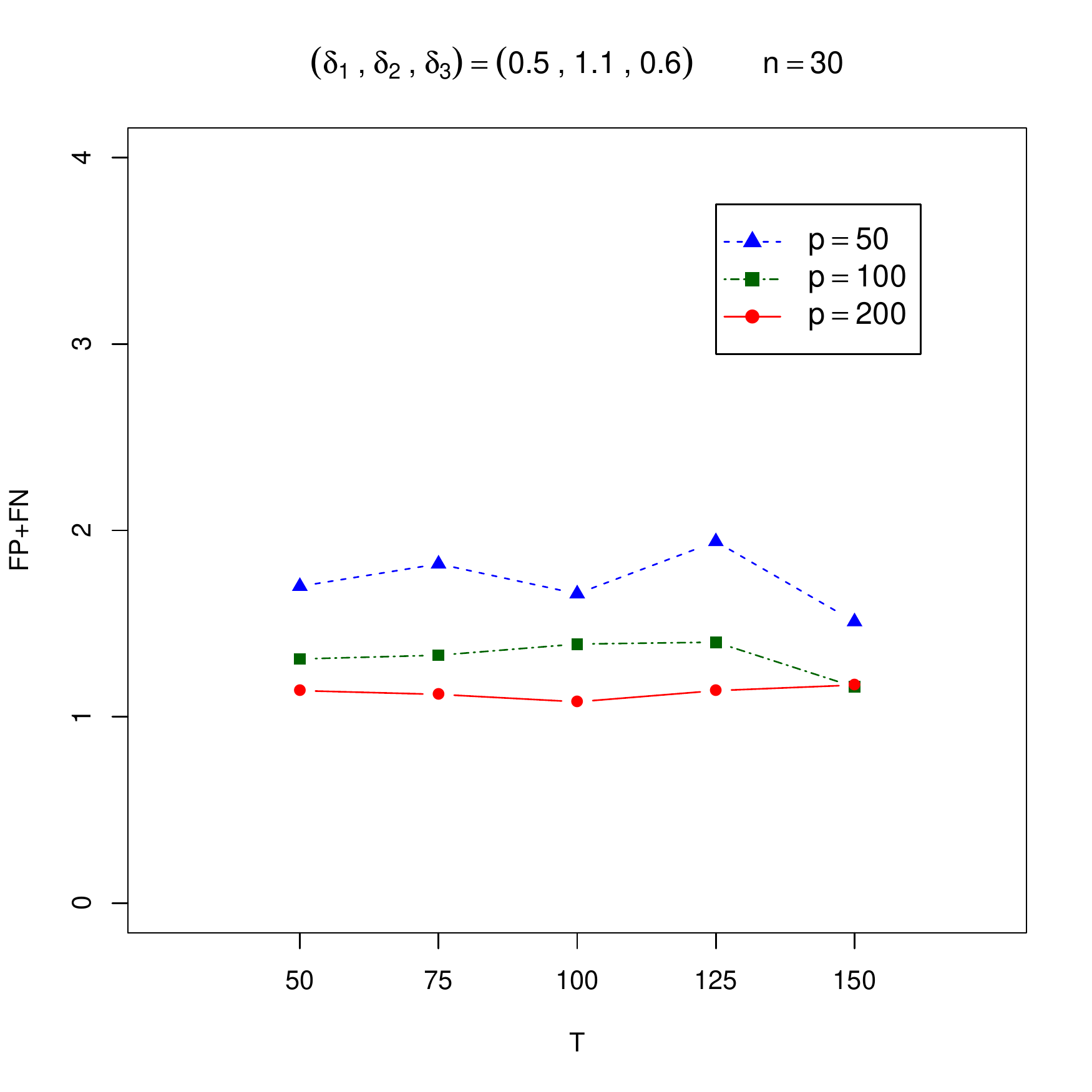}
\includegraphics[width=0.4\textwidth,height=0.4\textwidth]{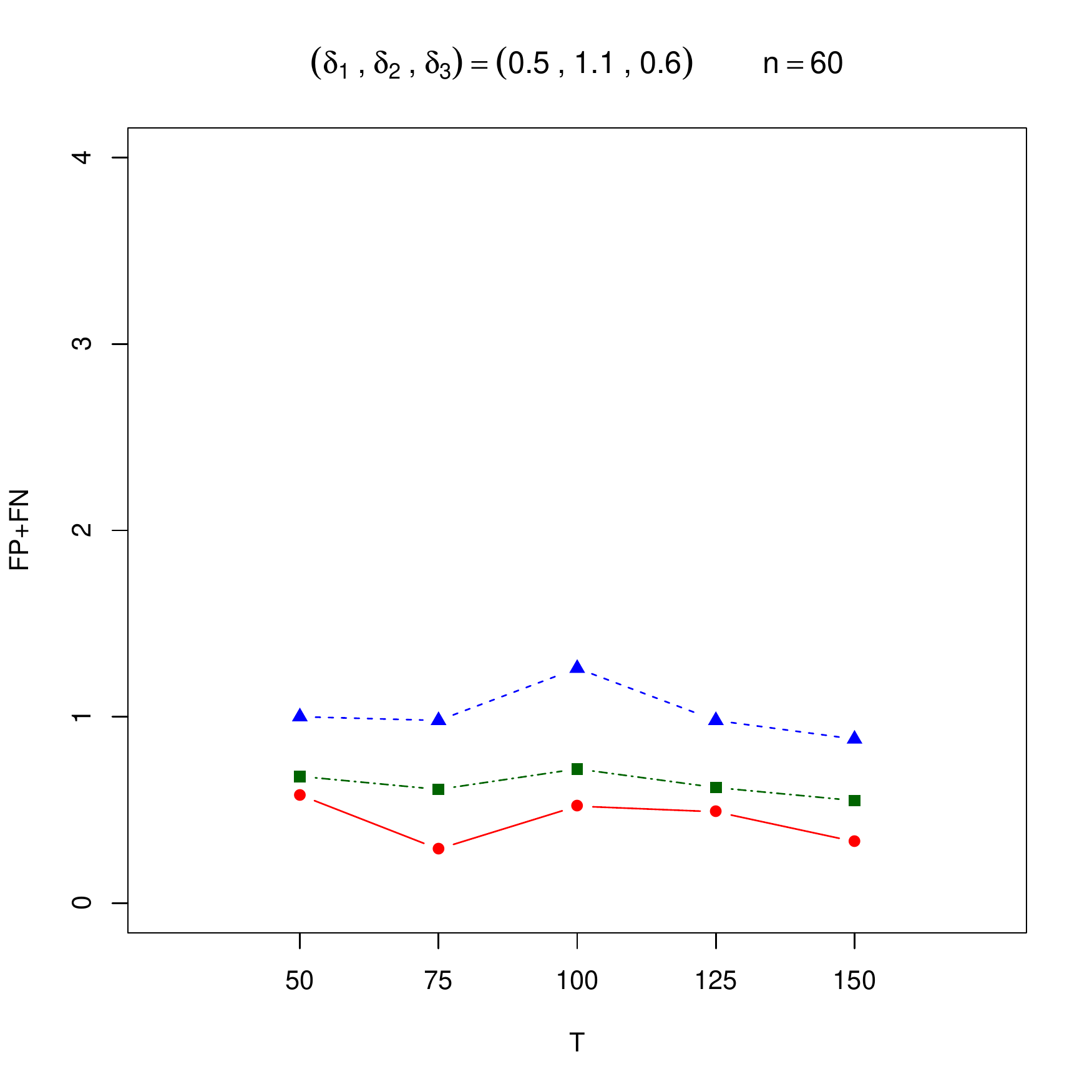}\\
\includegraphics[width=0.4\textwidth,height=0.4\textwidth]{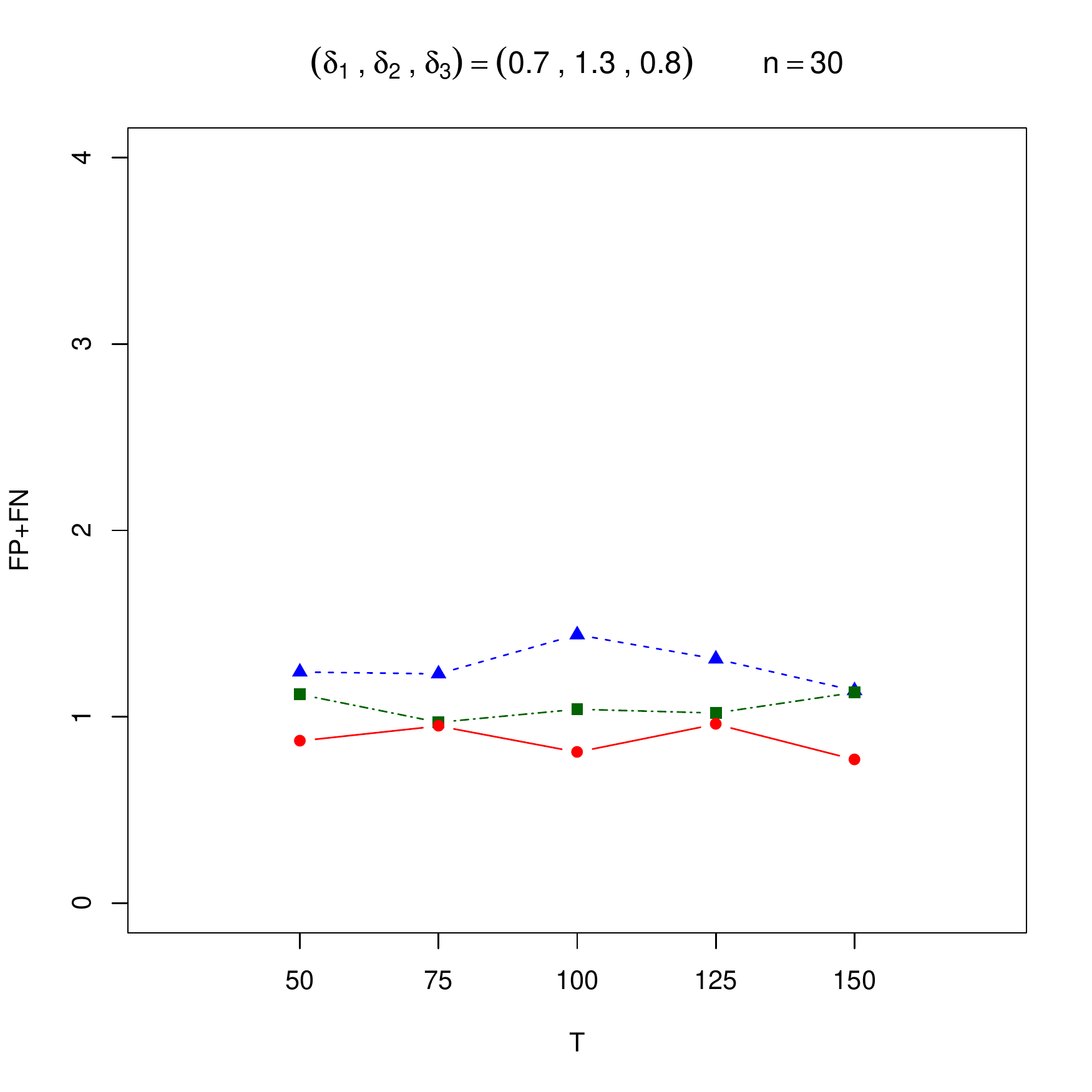}
\includegraphics[width=0.4\textwidth,height=0.4\textwidth]{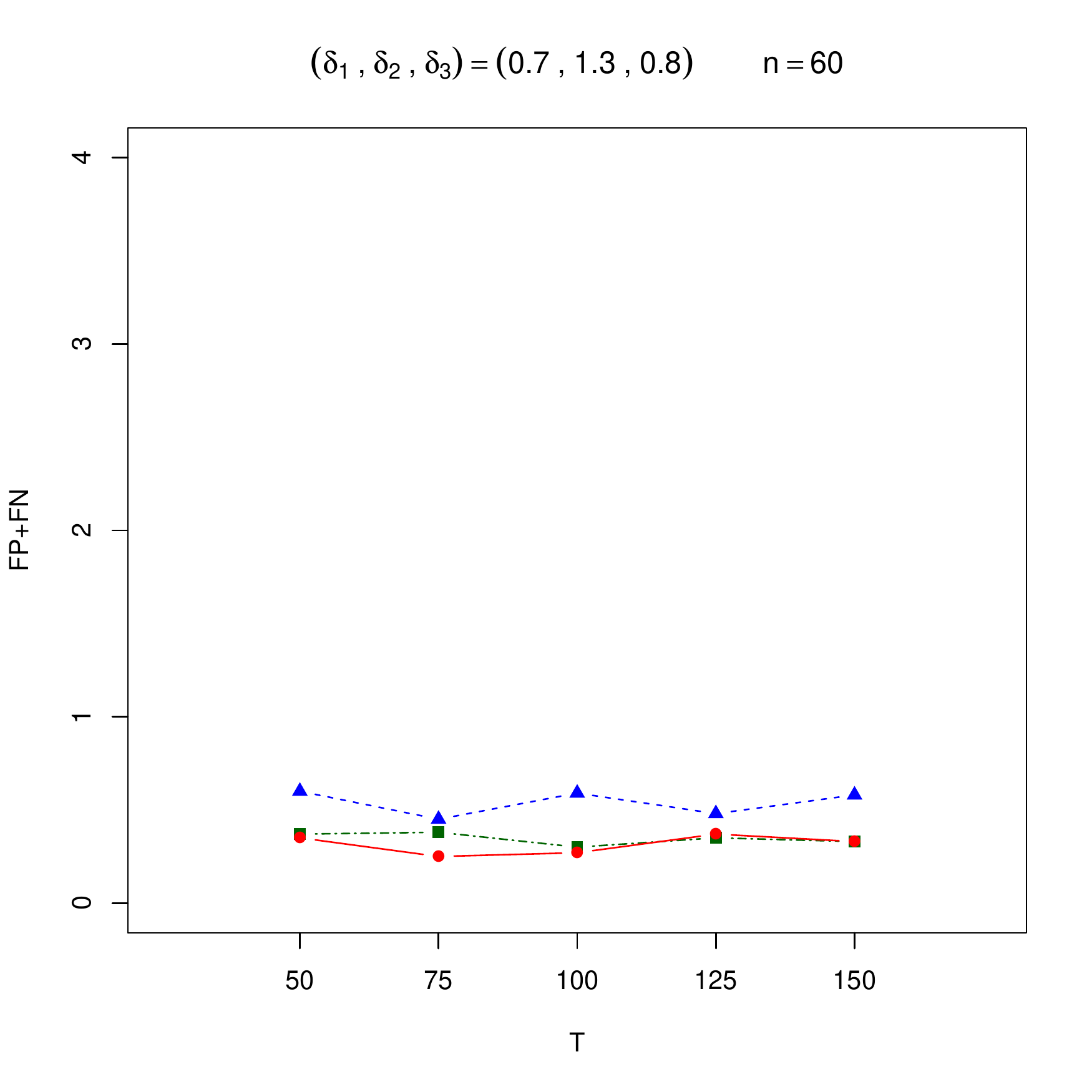} 
\caption{The average FP+FN under the mixture model with different combinations of signal strength $\delta$, dimension $p$, time $T$ and sample size $n$. The total number of change-points are set to be $2$. }
\label{fig4}
\end{center}
\end{figure}

Based on the same setup, we also conducted simulation experiment to evaluate performance of the proposed change-point identification procedure under the mixture model (\ref{mixture-of-means-simu}). 
The accuracy of the procedure is measured by the sum of FP and FN, which is illustrated in Figure \ref{fig4}. We observe that the patterns are similar to those reported in Figure \ref{fig2}. As $n$,
$p$ and $T$, or $(\delta_1,\delta_2, \delta_3$) increased, the FP+FN decreased. Specially, it was close to 0 when $p=200$, $n=60$ and $(\delta_1=0.7,\delta_2=1.3, \delta_3=0.8)$, showing that procedure is consistent under the mixture model.

\section{Real Data Analysis}

Recent studies suggest that the parahippocampal region of the brain activates more significantly to images with spatial structures than others without such structures (Epstein and Kanwisher, 1998; Henderson et al., 2007). An experiment was conducted to investigate the functions of such region in scene processing.  
During the experiment, fourteen students in Michigan State University were presented alternatively with six sets of 
scene images and six sets of object images. The order of presenting the images follows ``sososososoos'' where `s'  and `o' represent a set of scene images and object images, respectively.  
The fMRI data were acquired by placing each brain into a 3T GE Sigma EXCITE scanner. After the data were preprocessed by shifting time difference, correcting rigid-body motion and removing trends 
(more detail can be found in Henderson et al., 2011), the resulted dataset consists of BOLD measurements of 33,866 voxels from $14$ subjects and at $192$ time points, 
which clearly is a ``large $p$, large $T$ and small $n$" case. 

\begin{figure}[t!]
\begin{center}
\includegraphics[width=0.95\textwidth,height=5cm]{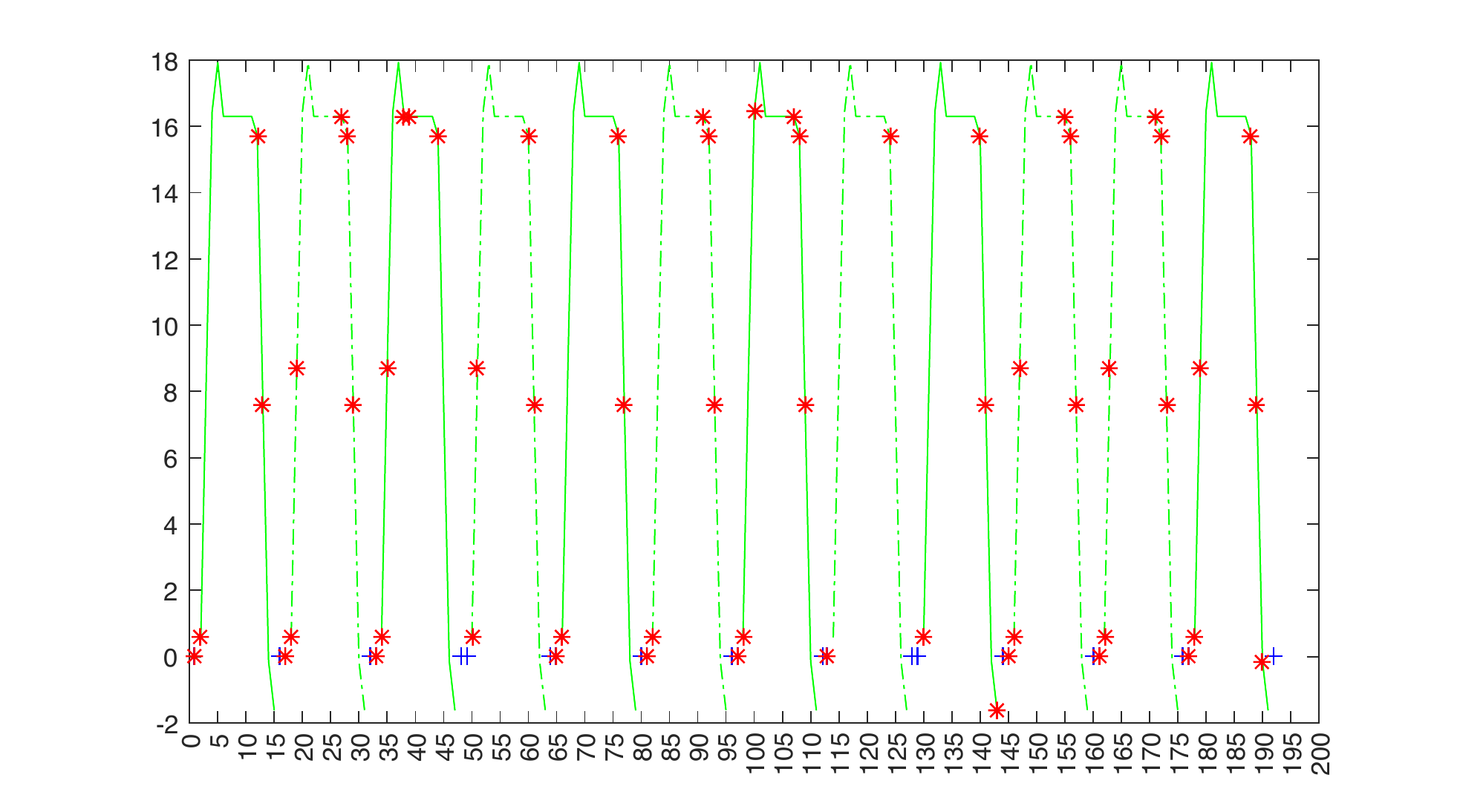}\vspace{-0.2cm}
\caption{The illustration of change-points identified by the proposed method. 
The green solid and dash curves, respectively, represent the expected BOLD responses to the scene and objective images. The x-values and y-values of the red stars marked on the curves, are the identified change-points and the corresponding BOLD responses. The blue plus signs represent the locations where subjects rest such that the BOLD responses are zero.  
Out of the 59 identified change-points, 58 are expected to have signal changes. 
}  
\label{cpts-fig}
\end{center}
\end{figure}


%

Let $X_{it}$ be a $p$-dim ($p=33,866$) random vector representing the fMRI image data for the $i$-th subject measured at time point $t$ ($i=1,\cdots, 14$ and $t=1,\cdots, 192$).
We first applied the testing procedure described in Section 4 to the dataset for testing the homogeneity of mean vectors, namely the hypothesis (\ref{general-hypo}). 
The test statistic $\tilde{\mathscr{M}}= 9.117$ with p-value less than $10^{-6}$, which indicates existence of change-points. After further implementing the proposed binary segmentation approach, 
we identified 59 change-points, which is not surprising because the large number of change-points arise from the time-altered scene and object images stimuli.      
To crosscheck the credibility of the identified change-points, we compared them with the predicted BOLD responses obtained from the convolution of the boxcar function with a gamma HRF function (Ashby, 2011). 
In Figure \ref{cpts-fig}, the green solid and the green dot dash curves following the order of presenting the images, are predicted BOLD responses to the scene images and object images, respectively. The x-values and y-values of the red stars marked on the curves, are the identified change-points and the corresponding BOLD responses. Based on the predicted BOLD response function, we found that 58 out of 59 identified change-points were expected to have signal changes. 
Keeping in mind that the proposed change-point detection and identification approach is nonparametric with no attempt to model neural activation, we have demonstrated that it has satisfactory performance for the fMRI data analysis. 

\begin{figure}[bt!]
\begin{center}
\includegraphics[width=0.26\textwidth,height=0.26\textwidth]{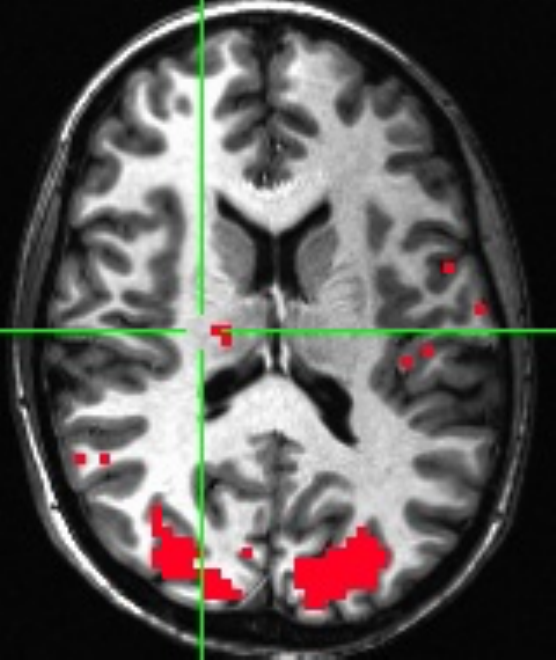}
\includegraphics[width=0.26\textwidth,height=0.26\textwidth]{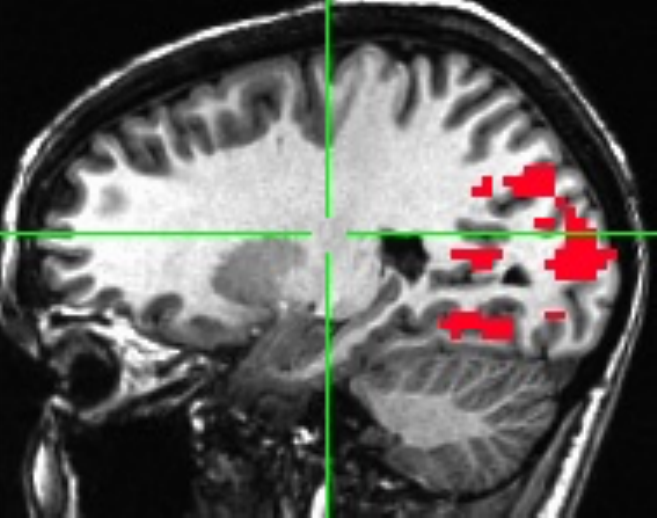}\\
\includegraphics[width=0.26\textwidth,height=0.26\textwidth]{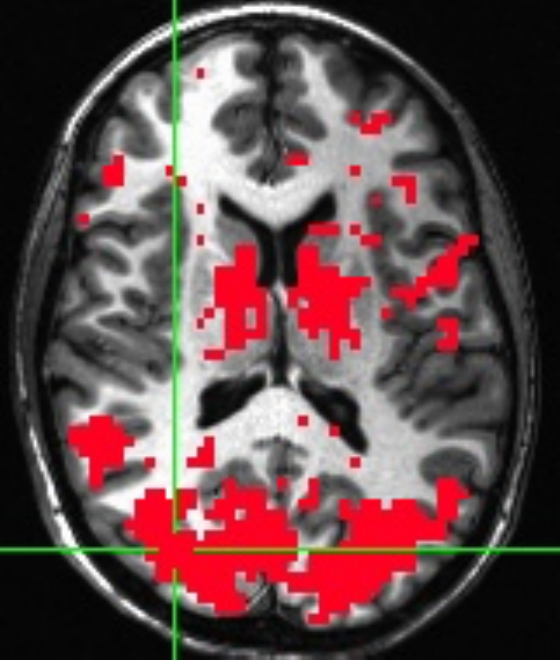}
\includegraphics[width=0.26\textwidth,height=0.26\textwidth]{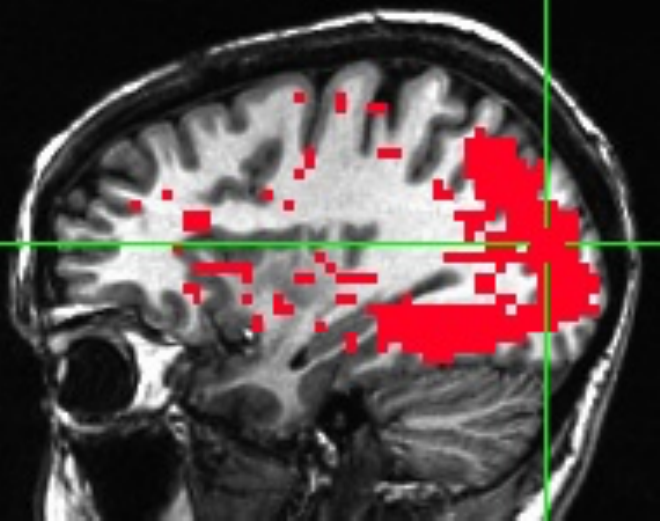}
\caption{Upper Panels: the activated brain regions at the 5th identified change-point (17th time point) where the object images were presented. Most of the significant changes (red areas) occurred at visual cortex areas. Lower Panels: the activated brain regions at  the 57th change-point (188th time point) where the scene images were presented. Most of the significant changes (red areas) occurred at both visual cortex and  parahippocampal areas.}
\label{significant-cpt5}
\end{center}
\end{figure}

To confirm that the parahippocampal region is selectively activated by the scenes over the objects, we compared the brain region activated by the scene images and with that activated by the object images. To do this, we let $X_{i \tau j}$ be the $j$-th component (voxel) of the random vector $X_{i \tau}$ for $i$-th subject at the change-point $\tau$ where $i=1, \cdots, 14$, $\tau=1, \cdots, 59$ and $j=1, \cdots, 33,866$. Similarly, let $X_{i \tau+1 j}$ be the $j$-th component of the random vector $X_{i \tau+1}$ after the change-point $\tau$. 
For each voxel ($j=1,\cdots, 33,866$), we computed the difference between two sample means $\bar{X}_{\tau j}$ and $\bar{X}_{\tau+1 j}$ and then conducted paired t-test for the significance of the mean difference before and after the change-point. 
Based on obtained p-values, we allocated the activated brain regions composed of all significant voxels after controlling the false discovery rate at $0.01$ (Storey, 2003). The results showed that the activated brain regions were quite similar across the same type of images, but significantly different between scene and object images. More specifically, the brain region activated by the scene images was located at both the visual cortex area and the parahippocampal area, whereas the region activated by the object images was only  located at the visual cortex area. 
Our findings are consistent with the results in Henderson et al. (2011). For illustration purpose, we only included pictures at two change-points in Figure \ref{significant-cpt5}. 

\section{Discussion}

Motivated by the real applications such as the fMRI studies, we consider the problem of testing the homogeneity of high dimensional mean vectors under the ``large $p$, large $T$ and small $n$'' paradigm.  We propose a new test statistic and establish its asymptotic distribution under mild conditions. One important feature of the proposed test is that it accommodates both temporal and spatial dependence. To the best of our knowledge, the temporal dependence has not been investigated in the literature of high dimensional MANOVA problems, so the proposed method has bridged this gap. When the null hypothesis is rejected, we further propose a procedure which is shown to be able to identify the change-points with probability converging to one. The rate of consistency of the change-point estimator is also established.

The proposed methods have also been generalized to a mixture model to allow heterogeneity among subjects. Numerical results demonstrate that the extension is promising and encouraging. Due to the space limitation, we will explore the theoretical results of the extension to the mixture model in a separate paper. Although the current article demonstrates the empirical performance of the proposed methods through the fMRI data analysis, they can be also applied to other high-dimensional longitudinal data.



\setcounter{equation}{0}
\def\theequation{A.\arabic{equation}}
\def\thesection{A}

\section*{\large Appendix: Technical Details}

In this Appendix, we provide proofs to the Theorems and Propositions in the paper.  Assume $\mu_t=0$ in (\ref{model0}) and (\ref{factor}). For any squared $m \times m$ matrix $A$ and $B$, the following results commonly 
used in Appendix can be derived: $\mbox{E}(X_{is}^{\prime}A X_{it})= \mbox{tr}(\Gamma_s^{\prime}A \Gamma_t),$
 and 
\begin{align}
\mbox{E}(X_{is}^{\prime}A X_{it}X_{is^*}^{\prime}B X_{it^*})&=\mbox{tr}(\Gamma_s^{\prime}A \Gamma_t)\mbox{tr}(\Gamma_{s^*}^{\prime}B \Gamma_{t^*})
+\mbox{tr}(\Gamma_s^{\prime}A \Gamma_t\Gamma_{s^*}^{\prime}B \Gamma_{t^*})\nonumber\\
&+\mbox{tr}(\Gamma_s^{\prime}A \Gamma_t\Gamma_{t^*}^{\prime}B^{\prime} \Gamma_{s^*})+(3+\Delta)\mbox{tr}(\Gamma_s^{\prime}A \Gamma_t \circ\Gamma_{s^*}^{\prime}B \Gamma_{t^*}),\label{com1}
\end{align}
where $A\circ B$ is the Hadamard product of $A$ and $B$.

\bigskip
\noindent{\bf A.1. Proof of Theorem 1.}
\bigskip

Theorem 1 can be established by the martingale central limit theorem. Toward this end, we first construct a martingale difference sequence. If we define $Y_{i s_a}=X_{i s_a}-\mu_{s_a}$, then
$\hat{M}_t-M_t=\sum_{i=1}^n M_{ti}, $
where 
\begin{align*}
M_{ti}=&\frac{2}{n(n-1)h(t)}\sum_{j=1}^{i-1}\big\{\sum_{s_1=1}^t \sum_{s_2=t+1}^T \sum_{a,b \in \{1,2\}} (-1)^{|a-b|}Y_{i s_a}^{\prime}Y_{j s_b}\big\}\\
&+\frac{2}{nh(t)} \sum_{s_1=1}^t \sum_{s_2=t+1}^T \sum_{a,b \in \{1,2\}} (-1)^{|a-b|} \mu^{\prime}_{s_a} Y_{i s_b}.
\end{align*} 
Let $\{\mathscr{F}_i, 1 \le i \le n\}$ be $\sigma$-fields generated by $\sigma\{\mathbb{Y}_1, \cdots, \mathbb{Y}_i \}$ where $\mathbb{Y}_i=\{Y_{i1}, \cdots, Y_{i T} \}^{\prime}$. 
Then it can be shown that ${\E}(M_{tk}|\mathscr{F}_{k-1})=0$ for $k=1, \cdots, n$. Therefore, $\{M_{ti}, 1\le i \le n\}$ is a martingale difference sequence with respect to $\sigma$-fields $\{\mathscr{F}_i, 1 \le i \le n\}$.  

Based on Lemmas 1 and 2 proved in the supplementary material, Theorem 1 can be proved using the martingale central limit theorem 
(Hall and Heyde, 1980).

\bigskip
\noindent{\bf A.2. Proof of Theorem 2.}
\bigskip

Note that the estimator $\widehat{\mbox{tr}(\Xi_{r_a s_c} \Xi_{r_b s_d}^{\prime})}$ in (\ref{variance-est}) is invariant by transforming $X_{it}$ to $X_{it}-\mu_{t}$ where $t=1, \cdots, \tau$. 
With loss of generality, we assume that $\mu_1=\mu_2=\cdots=\mu_T=0$. First, 
\begin{align*}
&\quad\mbox{E}\big\{\widehat{\mbox{tr}(\Xi_{r_a s_c} \Xi_{r_b s_d}^{\prime})} \big\}\\
&=\mbox{E}( X_{i r_a}^{\prime} X_{j r_b} X_{i s_c}^{\prime} X_{j s_d}) -\mbox{E} (X_{i r_a}^{\prime} X_{j r_b} X_{i s_c}^{\prime} X_{k s_d})\nonumber\\
&\quad -\mbox{E} (X_{i r_a}^{\prime} X_{j r_b} X_{k s_c}^{\prime} X_{j s_d})+\mbox{E}( X_{i r_a}^{\prime} X_{j r_b} X_{k s_c}^{\prime} X_{l s_d}) 
= \mbox{tr}(\Xi_{r_a s_c} \Xi_{r_b s_d}^{\prime}). \nonumber
\end{align*}
This shows that $\mbox{E}(\hat{\sigma}_{nt, 0}^2)=\sigma_{nt,0}^2$. Therefore, to prove Theorem 2, we only need to show that
$\mbox{Var}(\hat{\sigma}_{nt, 0}^2)/\sigma_{nt,0}^4 \to 0.$

For convenience, we denote the summation $\sum_{r_1=1}^t\sum_{r_2=t+1}^T\sum_{s_1=1}^t\sum_{s_2=t+1}^T$ by $\sum_{r_1, r_2, s_1, s_2}$. 
Define the right hand side of  ``$=$'' in (\ref{variance-est}) as $B_1+B_2+B_3+B_4$, and accordingly, 
\begin{eqnarray}
\hat{\sigma}_{nt,0}^2&=&\frac{2}{h^{2}(t)n(n-1)}\sum_{r_1, r_2, s_1, s_2} \sum_{a,b,c,d \in \{1,2\}}(-1)^{|a-b|+|c-d|} (B_1+B_2+B_3+B_4)\nonumber\\
&\equiv& \hat{\sigma}_{nt,0}^{2 (1)}+ \hat{\sigma}_{nt,0}^{2 (2)}+ \hat{\sigma}_{nt,0}^{2 (3)}+ \hat{\sigma}_{nt,0}^{2 (4)}.\nonumber
\end{eqnarray}
Therefore, we only need to show that $\mbox{Var}(\hat{\sigma}_{nt, 0}^{2 (i)})/ \sigma_{nt,0}^4 \to 0$ for $i=1,2,3$ and $4$ respectively. Toward this end, we first show that $\mbox{Var}(\hat{\sigma}_{nt, 0}^{2 (1)} )/ \sigma_{nt,0}^4 \to 0$ as follows.
\begin{align}
&\mbox{Var}(\hat{\sigma}_{nt, 0}^{2 (1)})\nn\\
=&\frac{4}{h^{4}(t)n^4(n-1)^4}\mbox{Var}\big\{ \sum_{r_1, r_2, s_1, s_2} \sum_{a,b,c,d \in \{1,2\}} (-1)^{|a-b|+|c-d|} \sum_{i \ne j}^n X_{i r_a}^{\prime} X_{j r_b} X_{i s_c}^{\prime} X_{j s_d}  \big\}\nonumber\\
=&\frac{4}{h^{4}(t)n^4(n-1)^4}\overline{\sum} \Big\{ \sum_{i \ne j, k\ne l}^n \mbox{E}(X_{i r_a}^{\prime} X_{j r_b} X_{i s_c}^{\prime} X_{j s_d}X_{k r^*_{a^*}}^{\prime} X_{l r^*_{b^*}} X_{k s^*_{c^*}}^{\prime} X_{l s^*_{d^*}})\nonumber\\
&\qquad\qquad\qquad
 -n^2(n-1)^2\mbox{tr}(\Gamma_{r_a}^{\prime} \Gamma_{r_b} \Gamma_{s_c}^{\prime} \Gamma_{s_d})\mbox{tr}(\Gamma_{r^*_{a^*}}^{\prime} \Gamma_{r^*_{b^*}} \Gamma_{s^*_{c^*}}^{\prime} \Gamma_{s^*_{d^*}})\Big\}, 
\label{Th2-1}
\end{align}
where $\overline{\sum}$ represents $\sum_{r_1, r_2, s_1, s_2} \sum_{a,b,c,d \in \{1,2\}}\sum_{r_1^*, r_2^*, s_1^*, s_2} \sum_{a^*,b^*,c^*,d^* \in \{1,2\}}$.

Now we evaluate $\sum_{i \ne j, k\ne l}^n \mbox{E}(X_{i r_a}^{\prime} X_{j r_b} X_{i s_c}^{\prime} X_{j s_d}X_{k r^*_{a^*}}^{\prime} X_{l r^*_{b^*}} X_{k s^*_{c^*}}^{\prime} X_{l s^*_{d^*}})$ with respect to different 
cases in the following.
First, if all indices are distinct, i.e., $i\ne j \ne k \ne l$. Using (\ref{com1}), we have
\begin{align*}
\sum_{i \ne j, k\ne l}^n &\mbox{E}(X_{i r_a}^{\prime} X_{j r_b} X_{i s_c}^{\prime} X_{j s_d}X_{k r^*_{a^*}}^{\prime} X_{l r^*_{b^*}} X_{k s^*_{c^*}}^{\prime} X_{l s^*_{d^*}})\\
&\asymp n^4\mbox{tr}(\Gamma_{r_a}^{\prime}\Gamma_{r_b}\Gamma_{s_d}^{\prime}\Gamma_{s_c})\mbox{tr}(\Gamma_{r^*_{a^*}}^{\prime}\Gamma_{r^*_{b^*}}\Gamma_{s^*_{d^*}}^{\prime}\Gamma_{s^*_{c^*}}).
\end{align*}
Next, if $(i=k)\ne j \ne l$, then by (\ref{com1}),
\begin{align*}
&\quad\sum_{i \ne j, k\ne l}^n \mbox{E}(X_{i r_a}^{\prime} X_{j r_b} X_{i s_c}^{\prime} X_{j s_d}X_{k r^*_{a^*}}^{\prime} X_{l r^*_{b^*}} X_{k s^*_{c^*}}^{\prime} X_{l s^*_{d^*}})\\
&\asymp n^3\big\{(3+\Delta) \mbox{tr}(\Gamma_{r_a}^{\prime}\Gamma_{r_b}\Gamma_{s_d}^{\prime}\Gamma_{s_c}\circ\Gamma_{r^*_{a^*}}^{\prime}\Gamma_{r^*_{b^*}}\Gamma_{s^*_{d^*}}^{\prime}\Gamma_{s^*_{c^*}})\\
&\quad+\mbox{tr}(\Gamma_{r_a}^{\prime}\Gamma_{r_b}\Gamma_{s_d}^{\prime}\Gamma_{s_c})\mbox{tr}(\Gamma_{r^*_{a^*}}^{\prime}\Gamma_{r^*_{b^*}}\Gamma_{s^*_{d^*}}^{\prime}\Gamma_{s^*_{c^*}})\\
&\quad+\mbox{tr}(\Gamma_{r_a}^{\prime}\Gamma_{r_b}\Gamma_{s_d}^{\prime}\Gamma_{s_c}\Gamma_{r^*_{a^*}}^{\prime}\Gamma_{r^*_{b^*}}\Gamma_{s^*_{d^*}}^{\prime}\Gamma_{s^*_{c^*}})+\mbox{tr}(\Gamma_{r_a}^{\prime}\Gamma_{r_b}\Gamma_{s_d}^{\prime}\Gamma_{s_c}\Gamma_{s^*_{c^*}}^{\prime}\Gamma_{s^*_{d^*}}\Gamma_{r^*_{b^*}}^{\prime}\Gamma_{r^*_{a^*}})\big\},\nonumber
\end{align*}
which is equal to other cases $(j=k)\ne i \ne l$, $(i=l)\ne j \ne k$ and $(j=l)\ne i \ne k$.
Finally, we consider the cases $(i=k) \ne (j=l)$ and $(i=l) \ne (j=k)$. For the case $(i=k) \ne (j=l)$,
\begin{align*}
&\quad\sum_{i \ne j, k\ne l}^n \mbox{E}(X_{i r_a}^{\prime} X_{j r_b} X_{i s_c}^{\prime} X_{j s_d}X_{k r^*_{a^*}}^{\prime} X_{l r^*_{b^*}} X_{k s^*_{c^*}}^{\prime} X_{l s^*_{d^*}})\\
\asymp& n^2\big\{3\mbox{tr}(\Gamma_{r_a}^{\prime}\Gamma_{r_b}\Gamma_{s_d}^{\prime}\Gamma_{s_c})\mbox{tr}(\Gamma_{r^*_{a^*}}^{\prime}\Gamma_{r^*_{b^*}}\Gamma_{s^*_{d^*}}^{\prime}\Gamma_{s^*_{c^*}})
+3Q_1+(3+\Delta)Q_2\\
&+3(3+\Delta) \mbox{tr}(\Gamma_{s_d}^{\prime}\Gamma_{s_c}\Gamma_{r_a}^{\prime}\Gamma_{r_b}\circ\Gamma_{s^*_{d^*}}^{\prime}\Gamma_{s^*_{c^*}}\Gamma_{r^*_{a^*}}^{\prime}\Gamma_{r^*_{b^*}})\\
&
+(3+\Delta)^2\sum_{\alpha \beta}(\Gamma_{r_a}^{\prime}\Gamma_{r_b})_{\alpha \beta}(\Gamma_{s_d}^{\prime}\Gamma_{s_c})_{\beta \alpha}(\Gamma_{r^*_{a^*}}^{\prime}\Gamma_{r^*_{b^*}})_{\alpha \beta}(\Gamma_{s_{d^*}}^{\prime}\Gamma_{s_{c^*}})_{\beta \alpha}\big\},\nonumber
\end{align*}
where $Q_1=\mbox{tr}(\Gamma_{s_d}^{\prime}\Gamma_{s_c}\Gamma_{r_a}^{\prime}\Gamma_{r_b}\Gamma_{s^*_{d^*}}^{\prime}\Gamma_{s^*_{c^*}}\Gamma_{r^*_{a^*}}^{\prime}\Gamma_{r^*_{b^*}})+\mbox{tr}(\Gamma_{s_d}^{\prime}\Gamma_{s_c}\Gamma_{r_a}^{\prime}\Gamma_{r_b}\Gamma_{r^*_{b^*}}^{\prime}\Gamma_{r^*_{a^*}}\Gamma_{s^*_{c^*}}^{\prime}\Gamma_{s^*_{d^*}})$ and $Q_2=\mbox{tr}(\Gamma_{r_a}^{\prime}\Gamma_{r_b}\Gamma_{s_d}^{\prime}\Gamma_{s_c}\circ\Gamma_{r^*_{a^*}}^{\prime}\Gamma_{r^*_{b^*}}\Gamma_{s^*_{d^*}}^{\prime}\Gamma_{s^*_{c^*}})+
\mbox{tr}(\Gamma_{r_a}^{\prime}\Gamma_{r_b}\Gamma_{r_{b^*}}^{\prime}\Gamma_{r_{a^*}}\circ\Gamma_{s_{d}}^{\prime}\Gamma_{s_{c}}\Gamma_{s^*_{c^*}}^{\prime}\Gamma_{s^*_{d^*}})\\+\mbox{tr}(\Gamma_{r_a}^{\prime}\Gamma_{r_b}\Gamma_{s_{d^*}}^{\prime}\Gamma_{s_{c^*}}\circ\Gamma_{r_{a^*}}^{\prime}\Gamma_{r_{b^*}}\Gamma_{s_{d}}^{\prime}\Gamma_{s_{c}})$.
It can be shown that the case $(j=l)\ne i \ne k$ is the as the case $(i=k) \ne (j=l)$.

Plugging all the above results into (\ref{Th2-1}), we have
\[
\mbox{Var}(\hat{\sigma}_{nt, 0}^{2 (1)})\asymp h^{-4}(t)n^{-5}\overline{\sum}\mbox{tr}(\Gamma_{r_b}^{\prime}\Gamma_{r_a}\Gamma_{s_c}^{\prime}\Gamma_{s_d}\Gamma_{s^*_{d^*}}^{\prime}\Gamma_{s^*_{c^*}}\Gamma_{r^*_{a^*}}^{\prime}\Gamma_{r^*_{b^*}})+h^{-4}(t)n^{-6}\mbox{tr}(A_{0t}^2).
\]
Following the same procedure, it can be also shown that $\mbox{Var}(\hat{\sigma}_{nt, 0}^{2 (j)})=o\{\mbox{Var}(\hat{\sigma}_{nt, 0}^{2 (1)})\}$ for $j=2, 3$ and $4$.
Then, using condition (C1), we have $\mbox{Var}(\hat{\sigma}_{nt, 0}^{2 (j)})/ \sigma_{nt,0}^4 \to 0$ for $j=1, 2, 3$ and $4$. This completes the proof of Theorem 2.

\bigskip
\noindent{\bf A.4. Proof of Theorem 3.}
\bigskip

First, we derive $\mbox{Cov}(\hat{M}_u, \hat{M}_v)$ for $u, v \in \{1, \cdots, T-1\}$ under $H_0$ of (\ref{Hypo}). Without loss of generality, we assume that $\mu_1=\mu_2=\cdots=\mu_T=0$. Recall that 
\begin{align}
\hat{M}_u&=\frac{1}{h(u)n(n-1)}\sum_{s_1=1}^u \sum_{s_2=u+1}^T \biggl\{{\sum_{i\ne j}^nX_{is_1}^{\prime}X_{js_1}}+{\sum_{i\ne j}^nX_{is_2}^{\prime}X_{js_2}}-2{\sum_{i\ne j}^nX_{is_1}^{\prime}X_{js_2}} \biggr\}, \nonumber\\
\hat{M}_v&=\frac{1}{h(v)n(n-1)}\sum_{s_1=1}^v \sum_{s_2=v+1}^T \biggl\{{\sum_{i\ne j}^nX_{is_1}^{\prime}X_{js_1}}+{\sum_{i\ne j}^nX_{is_2}^{\prime}X_{js_2}}-2{\sum_{i\ne j}^nX_{is_1}^{\prime}X_{js_2}} \biggr\}. \nonumber
\end{align}
Following similar derivations for the variance of $\hat{M}_t$ in the proof of Proposition 1 in the supplementary material, we can derive that
\begin{align*}
\mbox{Cov}(\hat{M}_u, \hat{M}_v)&=\frac{2}{h(u) h(v) n(n-1)}\sum_{r_1=1}^u\sum_{r_2=u+1}^T\sum_{s_1=1}^v\sum_{s_2=v+1}^T \\
&\times \sum_{a,b,c,d \in \{1,2\}}(-1)^{|a-b|+|c-d|}\mbox{tr}(\Xi_{r_a s_c}\Xi^{\prime}_{r_b s_d}).
\end{align*}

Next, we show that $\{\hat{M}_t\}_{t=1}^{T-1}$ follow a joint multivariate normal distribution when $T$ is fixed. According to the Cramer-word device, we only need to show that for any non-zero constant vector 
$a=(a_1, \cdots, a_{T-1})^{\prime}$, $\sum_{t=1}^{T-1} a_t \hat{M}_t$ is asymptotically normal under $H_0$ of (\ref{Hypo}). Toward this end, we note that 
$\mbox{Var}(\sum_{t=1}^{T-1} a_t \hat{M}_t)=\sum_{u=1}^{T-1} \sum_{v=1}^{T-1} a_u a_v \mbox{Cov}(\hat{M}_u, \hat{M}_v)$. Then we only need to show that 
$\sum_{t=1}^{T-1} a_t \hat{M}_t /\sqrt{\mbox{Var}(\sum_{t=1}^{T-1} a_t \hat{M}_t)} \xrightarrow{d} N(0, 1)$, which can be proved by the martingale central limit theorem. 
Since the proof is very similar to that of Theorem 1, we omit it. With the joint normality of $\{\hat{M}_t\}_{t=1}^{T-1}$, the distribution of $\hat{\mathscr{M}}\to \max_{1\leq t\leq T-1}Z_t$ 
can be established by the continuous mapping theorem. 

To establish the asymptotic distribution of $\hat{\mathscr{M}}$ for $T$ diverging case, we need to show that under $H_0$, $\max_{1\leq t\leq T-1}\sigma_{nt}^{-1}\hat{M}_t$ 
converges to $\max_{1\leq t\leq T-1} Z_t$ where $Z_t$ is a Gaussian process with mean $0$ and covariance $\Sigma_Z$.
To this end, we need to show (i) the joint asymptotic normality of $(\sigma_{nt_1}^{-1}\hat{M}_{t_1},\cdots, \sigma_{nt_d}^{-1}\hat{M}_{t_d})^{\prime}$ for $t_1<t_2<\cdots<t_d$. (ii) the tightness of 
$\max_{1\leq t\leq T-1}\sigma_{nt}^{-1}\hat{M}_t$. The proof of (i) is the similar to the proof of the joint asymptotic normality under finite $T$ case. We need to prove (ii).


To prove (ii), let $W_n(s_1,s_2)=\sum_{a,b\in\{1,2\}}(-1)^{|a-b|}\{n(n-1)\}^{-1}\sum_{i\neq j} X_{is_a}^{\prime}X_{js_b}$ and the first order projection 
as $W_{n1}(s_1)=\{n(n-1)\}^{-1}\sum_{i\neq j} X_{is_1}^{\prime}X_{js_1}$. Then we have the following Hoeffding-type decomposition for $\hat{M}_t$,
\begin{align*}
\hat{M}_t=\sum_{s_1=1}^t\sum_{s_2=t+1}^Tg_n(s_1,s_2)+\sum_{s_1=1}^t\sum_{s_2=t+1}^T\{W_{n1}(s_1)+W_{n2}(s_2)\}:=\hat{M}_t^{(1)}+\hat{M}_t^{(2)},
\end{align*}
where $g_n(s_1,s_2)=W_n(s_1,s_2)-W_{n1}(s_1)-W_{n2}(s_2)$. The covariance between $\hat{M}_t^{(1)}$ and $\hat{M}_t^{(2)}$ is 0. First, we compute the variances of $\hat{M}_t^{(2)}$ under the
the null hypothesis $H_0$. We first write $\hat{M}_t^{(2)}=(T-t)\sum_{s_1=1}^tW_{n1}(s_1)+t\sum_{s_2=t+1}^TW_{n2}(s_2):=\hat{M}_t^{(21)}+\hat{M}_t^{(22)}$.  Then we have
\begin{align*}
\mbox{Var}(\hat{M}_t^{(21)})=\frac{2(T-t)^2}{n(n-1)}\sum_{s_1=1}^t\sum_{r_1=1}^t \mbox{tr}(\Xi_{s_1 r_1}\Xi_{s_1r_1}^\prime)
\end{align*}
Similarly, we have
\begin{align*}
\mbox{Var}(\hat{M}_t^{(22)})=\frac{2t^2}{n(n-1)}\sum_{s_2=t+1}^T\sum_{r_2=t+1}^T \mbox{tr}(\Xi_{s_2 r_2}\Xi_{s_2r_2}^\prime).
\end{align*}
In addition, the covariance between $\hat{M}_t^{(21)}$ and $\hat{M}_t^{(22)}$ is,
\begin{align*}
\mbox{Cov}(\hat{M}_t^{(21)}, \hat{M}_t^{(22)})=\frac{2t(T-t)}{n(n-1)}\sum_{s_1=1}^t\sum_{s_2=t+1}^T\mbox{tr}(\Xi_{s_1 s_2}\Xi_{s_1s_2}^\prime).
\end{align*}
In summary, the variance for $\hat{M}_t^{(2)}$ is
\begin{align*}
\mbox{Var}(\hat{M}_t^{(2)})&=\frac{2}{n(n-1)}\sum_{s_1,r_1=1}^t \sum_{s_2,r_2=t+1}^T\{\mbox{tr}(\Xi_{s_1 r_1}\Xi_{s_1r_1}^\prime)+\mbox{tr}(\Xi_{s_2 r_2}\Xi_{s_2r_2}^\prime)\\
&\quad+2\mbox{tr}(\Xi_{s_1 s_2}\Xi_{s_1s_2}^\prime)\}.
\end{align*}
Moreover, we have
\begin{align*}
\mbox{Var}(\hat{M}_t^{(1)})&=\frac{4}{n(n-1)}\sum_{s_1=1}^t\sum_{s_2=t+1}^T\{\mbox{tr}(\Sigma_{s_1}\Sigma_{s_2})+\mbox{tr}(\Xi_{s_2 s_1}\Xi_{s_2s_1})\}\\
&\quad+\frac{4}{n(n-1)}\sum_{s_1\neq r_1=1}^t \sum_{s_2\neq r_2=t+1}^T\{\mbox{tr}(\Xi_{s_1 r_1}\Xi_{s_2r_2}^\prime)+\mbox{tr}(\Xi_{s_2 r_1}\Xi_{s_1r_2}^\prime)\}.
\end{align*}
According to the condition (C2), $\mbox{tr}(\Xi_{s_1 r_1}\Xi_{s_1r_1}^\prime)\asymp \phi(|s_1-r_1|)\mbox{tr}(\Sigma_{s_1}\Sigma_{r_1})$ and $\sum_{k=1}^T\phi^{1/2}(k)<\infty$. Under the null hypothesis $H_0$, we have
\begin{align*}
&\mbox{Var}(\hat{M}_t^{(2)})\\
&\asymp \frac{2\mbox{tr}(\Sigma^2)}{n(n-1)}\sum_{s_1,r_1=1}^t \sum_{s_2,r_2=t+1}^T\{\phi(|s_1-r_1|)+\phi(|s_2-r_2|)+2\phi(|s_1-s_2|)\}\\
&\asymp \frac{2\mbox{tr}(\Sigma^2)}{n(n-1)}\{(T-t)^2t+t^2(T-t)\}.
\end{align*}
 On the other hand, we notice that the first term of $\mbox{Var}(\hat{M}_t^{(1)})$ has the same order as $t(T-t)\mbox{tr}(\Sigma^2)/\{n(n-1)\}$.
 Using the Cauchy-Schwarz inequality and under $H_0$, we have 
 $$
 \mbox{tr}^2(\Xi_{s_1 r_1}\Xi_{s_2r_2}^\prime)\leq \mbox{tr}(\Xi_{s_1 r_1}\Xi_{s_1r_1}^\prime)\mbox{tr}(\Xi_{s_2 r_2}\Xi_{s_2r_2}^\prime)\asymp 
 \phi(|s_1-r_1|)\phi(|s_2-r_2|)\mbox{tr}^2(\Sigma^2).
 $$
 Therefore, using the condition $\sum_{k=1}^T\phi^{1/2}(k)<\infty$, the second term in $\mbox{Var}(\hat{M}_t^{(1)})$ is also of order $t(T-t)\mbox{tr}(\Sigma^2)/\{n(n-1)\}$. In summary, $\hat{M}_t^{(1)}$
 is a small order of $\hat{M}_t^{(2)}$.  This also implies that $\sigma_{nt}^2=\mbox{Var}(\hat{M}_t^{(2)})\{1+o(1)\}$. 
 
 Consider $t=[T\nu]$ for $\nu=j/T\in(0,1)$ with $j=1,\cdots, T-1$. Based on the above results, to show the tightness of $\max_{1\leq t\leq T-1}\sigma_{nt}^{-1}\hat{M}_t$ is equivalent to show the tightness
 of $G_n(\nu)$ where 
 $$G_n(\nu)=T^{-3/2}n^{-1}\mbox{tr}^{-1/2}(\Sigma^2)(\hat{M}_{[T\nu]}^{(1)}+\hat{M}_{[T\nu]}^{(2)}):=G_n^{(1)}(\nu)+G_n^{(2)}(\nu).$$
 
 We first show the tightness of $G_n^{(1)}(\nu)$. To this end, we first note that, for $1>\eta>\nu>0$, 
 \begin{align*}
 &E\Big\{|G_n^{(1)}(\nu)-G_n^{(1)}(\eta)|^2\Big\}\\
 &=\frac{1}{T^{3}n^{2}\mbox{tr}(\Sigma^2)}E\Big\{\Big|\sum_{s_1=1}^{[T\nu]}\sum_{s_2=[T\nu]+1}^{[T\eta]}g_n(s_1,s_2)-\sum_{s_1=[T\nu]+1}^{[T\eta]}\sum_{s_2=[T\eta]+1}^{T}g_n(s_1,s_2)\Big|^2\Big\}\\
 &\leq CT^{-3} \{[T\nu]([T\eta]-[T\nu])+(T-[T\eta])([T\eta]-[T\nu])\}\leq C(\eta-\nu)/T.
\end{align*}
Applying the above inequality with $\nu=k/T$ and $\eta=m/T$ for $0\leq k\leq m< T$ for integers $k, m$ and $T$ and using Chebyshev's inequality, we have, for any $\epsilon>0$,
\small
 \begin{align*}
 P\Big(\Big|G_n^{(1)}(k/T)-G_n^{(1)}(m/T)\Big|\geq \epsilon\Big)&\leq E\Big\{|G_n^{(1)}(k/T)-G_n^{(1)}(m/T)|^2\Big\}/\epsilon^2\\
&\leq C(m-k)/(\epsilon T)^2\leq (C/\epsilon^2)(m-k)^{1+\alpha}/T^{2-\alpha},
\end{align*}
\normalsize
where $0<\alpha<1/2$. Now if we define $\xi_i=G_n^{(1)}(i/T)-G_n^{(1)}((i-1)/T)$ for $i=1,\cdots, T-1$. Then $G_n^{(1)}(i/T)$ is equal to the partial sum of $\xi_i$, namely $S_i=\xi_1+\cdots+\xi_i=G_n^{(1)}(i/T)$. Here $S_0=0$.
Then we have 
$$P(|S_m-S_k|\geq \epsilon)\leq (1/\epsilon^2)\{C^{1/(1+\alpha)}(m-k)/T^{(2-\alpha)/(1+\alpha)}\}^{1+\alpha}.$$
Then using Theorem 10.2 in Billingsley (1999), we conclude the following
$$
P(\max_{1\leq i\leq T}|S_i|\geq \epsilon)\leq (KC/\epsilon^2)\{T/T^{(2-\alpha)/(1+\alpha)}\}^{1+\alpha}\leq (KC/\epsilon^2)T^{-1+2\alpha}.
$$
The right hand side of the above inequality goes to 0 as $T\to\infty$ because $\alpha<1/2$. Based on the relationship between $S_i$ and $G_n^{(1)}(i/T)$, we have shown the
tightness of $G_n^{(1)}(\nu)$.

Next, we consider the tightness of $G_n^{(2)}(\nu)$. Recall that 
\begin{align*}
G_n^{(2)}(\nu)&=T^{-3/2}n^{-1}\mbox{tr}^{-1/2}(\Sigma^2)\sum_{s_1=1}^{[T\nu]}\sum_{s_2=[T\nu]+1}^T\{W_{n1}(s_1)+W_{n2}(s_2)\}\\
&=T^{-3/2}n^{-1}\mbox{tr}^{-1/2}(\Sigma^2)(T-[T\nu])\sum_{s_1=1}^{[T\nu]}W_{n1}(s_1)\\
&\quad+T^{-3/2}n^{-1}\mbox{tr}^{-1/2}(\Sigma^2)[T\nu]\sum_{s_2=[T\nu]+1}^{T}W_{n2}(s_2):=G_n^{(21)}(\nu)+G_n^{(22)}(\nu).
\end{align*}

It is enough to show the tightness of $G_n^{(21)}(\nu)$, since the tightness of $G_n^{(22)}(\nu)$ is similar.
Let $h(i,j)=T^{-1/2}\sum_{s_1=[T\nu]+1}^{[T\eta]} (X_{is_1}-\mu)^{\prime}(X_{js_1}-\mu).$ Then, we have the following
\begin{align*}
G_n^{(21)}(\eta)-G_n^{(21)}(\nu)&=T^{-1/2}n^{-1}\mbox{tr}^{-1/2}(\Sigma^2)\sum_{s_1=[T\nu]+1}^{[T\eta]} \frac{1}{n(n-1)}\sum_{i\neq j} X_{is_1}^{\prime}X_{js_1}\\
&=\frac{1}{\sqrt{n(n-1)}\mbox{tr}(\Sigma^2)}\sum_{i\neq j} h(i,j).
\end{align*}
First, note that
\begin{align*}
&\{G_n^{(21)}(\eta)-G_n^{(21)}(\nu)\}^2\\
&=\frac{2}{n(n-1)\mbox{tr}(\Sigma^2)}\sum_{i\neq j} h^2(i,j)+\frac{4}{n(n-1)\mbox{tr}(\Sigma^2)}\sum_{i\neq j\neq k} h(i,j)h(i,k)\\
&\quad+\frac{1}{n(n-1)\mbox{tr}(\Sigma^2)}\sum_{i\neq j\neq k\neq l} h(i,j)h(k,l).
\end{align*}
Then, we have the following
\begin{align*}
E[\{G_n^{(21)}(\eta)-G_n^{(21)}(\nu)\}^4]&\leq E\Big[\frac{8}{n^2(n-1)^2\mbox{tr}^2(\Sigma^2)}\big\{\sum_{i\neq j} h^2(i,j)\big\}^2\Big]\\
&\quad+E\Big[\frac{32}{n^2(n-1)^2\mbox{tr}^2(\Sigma^2)}\big\{\sum_{i\neq j\neq k} h(i,j)h(i,k)\big\}^2\Big]\\
&\quad+E\Big[\frac{2}{n^2(n-1)^2\mbox{tr}^2(\Sigma^2)}\big\{\sum_{i\neq j\neq k\neq l} h(i,j)h(k,l)\big\}^2\Big]\\
&:=I_1+I_2+I_3.
\end{align*}
 
First, we consider $I_1$ in the above expression.
\begin{align*}
I_1&=E\Big[\frac{8}{n^2(n-1)^2\mbox{tr}^2(\Sigma^2)}\sum_{i\neq j}\sum_{i_i\neq j_1} h^2(i,j)h^2(i_1,j_1)\Big]\\
&=E\Big[\frac{16}{n^2(n-1)^2\mbox{tr}^2(\Sigma^2)}\sum_{i\neq j} h^4(i,j)\Big]\\
&\quad+E\Big[\frac{32}{n^2(n-1)^2\mbox{tr}^2(\Sigma^2)}\sum_{i\neq j\neq k} h^2(i,j)h^2(i,k)\Big]\\
&\quad+E\Big[\frac{8}{n^2(n-1)^2\mbox{tr}^2(\Sigma^2)}\sum_{i\neq j\neq i_i\neq j_1} h^2(i,j)h^2(i_1,j_1)\Big]:=I_{11}+I_{12}+I_{13}.
\end{align*}
We see that 
\begin{align*}
I_{13}&\asymp \frac{C}{T^2\mbox{tr}^2(\Sigma^2)}\Big\{\sum_{s_1=[T\nu]+1}^{[T\eta]}\sum_{r_1=[T\nu]+1}^{[T\eta]}\mbox{tr}(\Xi_{s_1r_1}\Xi_{s_1r_1}^\prime)\Big\}^2\asymp \frac{C}{T^2}\big\{[T\eta]-[T\nu]\big\}^2.
\end{align*}
After some calculation, we obtain that
\begin{align*}
I_{11}&=\frac{C}{n(n-1)T^2\mbox{tr}^2(\Sigma^2)}\Big[\Big\{\sum_{s_1=[T\nu]+1}^{[T\eta]}\sum_{r_1=[T\nu]+1}^{[T\eta]}\mbox{tr}(\Xi_{s_1r_1}\Xi_{s_1r_1}^\prime)\Big\}^2\\
&\quad+\sum_{s_1=[T\nu]+1}^{[T\eta]}\sum_{r_1=[T\nu]+1}^{[T\eta]}\sum_{u_1=[T\nu]+1}^{[T\eta]}\sum_{v_1=[T\nu]+1}^{[T\eta]} \mbox{tr}(\Xi_{r_1s_1}\Xi_{s_1v_1}\Xi_{v_1u_1}\Xi_{u_1r_1})\Big]=o(I_{13}).
\end{align*}
Similarly, it can be shown that $I_{12}=o(I_{13})$. In summary, $I_1\leq C\big\{[T\eta]-[T\nu]\big\}^2/T^2.$

Now, we check $I_2$. We have the following
\begin{align*}
I_2&=E\Big[\frac{64}{n^2(n-1)^2\mbox{tr}^2(\Sigma^2)}\sum_{i\neq i_1\neq j\neq k}h(i,j)h(i,k)h(i_1,j)h(i_1,k)\Big]\\
&\quad+E\Big[\frac{64}{n^2(n-1)^2\mbox{tr}^2(\Sigma^2)}\sum_{i\neq j\neq k}h(i,j)h(i,k)h(i,j)h(i,k)\Big]:=I_{21}+I_{22}.
\end{align*}
It can be seen that
\begin{align*}
I_{21}&\leq \frac{C}{\mbox{tr}^2(\Sigma^2)}E\Big[h(i,j)h(i,k)h(i_1,j)h(i_1,k)\Big]\\
&=\frac{C}{T^2\mbox{tr}^2(\Sigma^2)}\sum_{s_1,r_1,u_1,v_1}\mbox{tr} (\Xi_{s_1r_1}\Xi_{r_1v_1}\Xi_{v_1u_1}\Xi_{u_1s_1}),
\end{align*}
which is a smaller order of $I_{13}$. 
For $I_{22}$, we have 
\small
\begin{align*}
I_{22}&=\frac{C}{n\mbox{tr}^2(\Sigma^2)}E\Big[h(i,j)h(i,k)h(i,j)h(i,k)\Big]\\
&=\frac{C}{nT^2\mbox{tr}^2(\Sigma^2)}\sum_{s_1,r_1,u_1,v_1}\Big\{\mbox{tr}(\Xi_{s_1u_1}\Xi_{s_1u_1}^\prime)\mbox{tr}(\Xi_{r_1v_1}\Xi_{r_1v_1}^\prime)
+\mbox{tr} (\Xi_{s_1u_1}\Xi_{u_1r_1}\Xi_{r_1v_1}\Xi_{v_1s_1})\Big\}.
\end{align*}
\normalsize
Therefore, $I_{22}$ is also a smaller order of $I_{13}$. In summary, $I_1$ is a smaller oder of $I_{13}$.

At last, let us consider $I_3$. After some calculation, we have the following
\begin{align*}
I_{3}&\asymp E\Big[\frac{C}{\mbox{tr}^2(\Sigma^2)}\{h^2(i,j)h^2(k,l)+h(i,j)h(k,l)h(i,k)h(j,l)\}\Big]\\
&=\frac{C}{T^2\mbox{tr}^2(\Sigma^2)}\Big\{\sum_{s_1=[T\nu]+1}^{[T\eta]}\sum_{r_1=[T\nu]+1}^{[T\eta]}\mbox{tr}(\Xi_{s_1r_1}\Xi_{s_1r_1}^\prime)\Big\}^2\\
&\quad+\frac{C}{T^2\mbox{tr}^2(\Sigma^2)}\sum_{s_1,r_1,u_1,v_1}\mbox{tr} (\Xi_{s_1r_1}\Xi_{r_1v_1}\Xi_{v_1u_1}\Xi_{u_1s_1}).
\end{align*}
Now it is clear that the first term in $I_3$ is of the same order as $I_{13}$ and the second term is of the same order as $I_{21}$. Therefore, $I_3\leq C\big\{[T\eta]-[T\nu]\big\}^2/T^2.$

Let $\nu=k/T$ and $\eta=m/T$ for $0\leq k\leq m< T$ for integers $k, m$ and $T$ and using the above bounds for the fourth moment of $|G_n^{(21)}(\eta)-G_n^{(21)}(\nu)|$, we have, for any $L>0$,
 \begin{align*}
 P\Big(\Big|G_n^{(21)}(k/T)-G_n^{(21)}(m/T)\Big|\geq L\Big)&\leq E\Big\{|G_n^{(21)}(k/T)-G_n^{(21)}(m/T)|^4\Big\}/L^4\\
&\leq (C/L^4)\{(m-k)/T\}^2.
\end{align*}
Applying Theorem 10.2 in Billingsley (1999) again, we have 
$$
P(\max_{1\leq i\leq T}|G_n^{(21)}(i/T)|\geq L)\leq KC/L^4.
$$
If $L$ is large enough, the above probability could be smaller than any $\epsilon>0$. Therefore, $\max_{1\leq i\leq T}|G_n^{(21)}(i/T)|$ is tight. Similarly, we can show the tightness of $\max_{1\leq i\leq T}|G_n^{(22)}(i/T)|$.
In summary, we have shown the tightness of $G_n^{(1)}(\nu)$ and $G_n^{(2)}(\nu)$. Hence, $G_n(\nu)$ is also tight. 
Combining (i) and (ii) together, we know that $\sigma_{nt}^{-1}\hat{M}_t$ converges to a Gaussian process with mean 0 and covariance $\Sigma_Z$.

Finally, applying Lemma 4 in the supplementary material,
we can show that the asymptotic distribution of $\max_{1\leq t\leq T-1}\sigma_{nt, 0}^{-1}\hat{M}_t$ is the desired Gumbel distribution.
This completes the proof of Theorem 3.

\bigskip
\noindent{\bf A.5. Proof of Theorem 4.}
\bigskip

Recall that $\sigma_{\max}=\max_{0< t/T<1}\max \{\sqrt{\mbox{tr}(A_{0t}^2)/h^2(t)}, \sqrt{n||A_{1t}||^2/h^2(t)}\}$ and $\delta=\|\mu_1-\mu_T\|^2$.
Given a constant $C$, we define a set $$K(C)=\{t: |t-\tau| > C T\mbox{log}^{1/2} T \sigma_{\max}/(n\delta), \quad 1 \le t \le T-1 \}.$$ To show Theorem 4, we first show that for any $\epsilon>0$, there exists a constant $C$ such that
\be
\mbox{P}\big\{|\hat{\tau}-\tau|> C T \mbox{log}^{1/2} T  \sigma_{\max}/(n\delta)\big\} < \epsilon. \label{consist}
\ee
Since the event $\{\hat{\tau} \in K(C)\}$ implies the event $\{\max_{t \in K(C)} \hat{M}_t > \hat{M}_{\tau} \}$, then it is enough to show that
\[
\mbox{P}\big(\max_{t \in K(C)} \hat{M}_t > \hat{M}_{\tau} \big) < \epsilon.
\]
Toward this end, we first derive the result based on the definition of $M_t$:
\[
M_t=\Big\{\frac{T-\tau}{T-t}I(1\le t \le \tau)
+\frac{\tau}{t}I(\tau<t \le T)\Big\}\delta,
\]
where $\delta=(\mu_{1}-\mu_{T})^{\prime}(\mu_{1}-\mu_{T})$. Specially, $M_t$ attains its maximum $\delta$ at $t=\tau$ since $1/(T-t)$ is an increasing function and $1/t$ is a decreasing function. As a result, by union sum inequality and letting $A(t, \tau |1, T)={1}/{(T-t)}I(1\le t \le \tau)+{1}/{t}I(\tau<t \le T)$, we have
\begin{align*}
\mbox{P}\biggl(\max_{t \in K(C)} \hat{M}_t > \hat{M}_{\tau} \biggr)&\le \sum_{t \in K(C)} \mbox{P}\biggl( \hat{M}_t-M_t+M_t-M_{\tau} > \hat{M}_{\tau}-M_{\tau} \biggr)\\
&\le\sum_{t \in K(C)} \mbox{P}\biggl\{\Big|\frac{\hat{M}_t-M_t}{\sigma_{nt}}\Big| > \frac{A(t, \tau |1, T)}{2}\frac{\delta}{\sigma_{\max}}|\tau-t|\biggr\} \\
&+\sum_{t \in K(C)} \mbox{P}\biggl\{\Big|\frac{\hat{M}_{\tau}-M_{\tau}}{\sigma_{n\tau}}\Big| > \frac{A(t, \tau |1, T)}{2}\frac{\delta}{\sigma_{\max}}|\tau-t|\biggr\}\\
&\le\sum_{t \in K(C)} \mbox{P}\biggl\{\Big|\frac{\hat{M}_t-M_t}{\sigma_{nt}}\Big| > \sqrt{C \mbox{log} T}\biggr\}\\
&+\sum_{t \in K(C)} \mbox{P}\biggl\{\Big|\frac{\hat{M}_{\tau}-M_{\tau}}{\sigma_{n\tau}}\Big|> \sqrt{C \mbox{log} T}\biggr\},
\end{align*}
where the result of $A(t, \tau|1, T)=O(1/T)$ has been used. 

Since $({\hat{M}_t-M_t})/{\sigma_{nt}} \sim \mbox{N}(0, 1)$, for a large $C$, 
\[
\sum_{t \in K(C)}\mbox{P}\biggl\{\Big|\frac{\hat{M}_t-M_t}{\sigma_{nt}}\Big| > \sqrt{C \mbox{log} T}\biggr\}=\sum_{t \in K(C)} C (\mbox{log} T)^{-1/2} T^{-C} 
\le \epsilon. 
\]
Similarly, we can show that 
\[
\sum_{t \in K(C)} \mbox{P}\biggl\{\Big|\frac{\hat{M}_{\tau}-M_{\tau}}{\sigma_{n\tau}}\Big|> \sqrt{C \mbox{log} T}\biggr\} \le \epsilon. 
\]
Hence, (\ref{consist}) is true, which implies that $\hat{\tau}-\tau=O_p\big\{T\mbox{log}^{1/2} T \sigma_{\max}/(n\delta)\big\}$.

Recall that ${\sigma}_{\max}=\max_{0< t/T<1}\max \{\sqrt{\mbox{tr}(A_{0t}^2)/h^2(t)}, \sqrt{n||A_{1t}||^2/h^2(t)}\}$ and the assumption 
$\mbox{tr}(\Xi_{s_1 r_1}\Xi_{s_1r_1}^\prime)\asymp \phi(|s_1-r_1|)\mbox{tr}(\Sigma_{s_1}\Sigma_{r_1})$ and $\sum_{k=1}^T\phi^{1/2}(k)<\infty$,
following the proofs in Theorem 3, we have $\mbox{tr}(A_{0t}^2)\asymp T^3\mbox{tr}(\Sigma^2)$. Thus we have $\mbox{tr}(A_{0t}^2)/h^2(t)\asymp \mbox{tr}(\Sigma^2)/T$.

For the second part in ${\sigma}_{\max}$, if $1\leq t\leq \tau$, we have 
\begin{align*}
\|A_{1t}\|^2=(\mu_1-\mu_T)^{\prime}\sum_{r_1,s_1=1}^t\sum_{r_2,s_2=t+1}^T(\Gamma_{r_1}-\Gamma_{r_2})(\Gamma_{s_1}-\Gamma_{s_2})^{\prime}(\mu_1-\mu_T).
\end{align*}
Using the assumption that $(\mu_1-\mu_T)^{\prime}\Xi_{r_1s_1}(\mu_1-\mu_T)\asymp \phi(|r_1-s_1|)(\mu_1-\mu_T)^{\prime}\Sigma(\mu_1-\mu_T)$, it can be checked that
$\|A_{1t}\|^2\asymp T^3 (\mu_1-\mu_T)^{\prime}\Sigma(\mu_1-\mu_T)$. In summary, we have
$$
\sigma_{\max}=\max \{\sqrt{\mbox{tr}(\Sigma^2)}, \sqrt{n(\mu_1-\mu_T)^{\prime}\Sigma(\mu_1-\mu_T)}\}/\sqrt{T}=v_{\max}/\sqrt{T}.
$$
This completes the proof of Theorem 4.

\bigskip
\noindent{\bf A.6. Proof of Theorem 5.}
\bigskip

To prove Theorem 5, we need the following Lemma \ref{lem3}, whose proof is presented in the supplementary material. 
The Lemma \ref{lem3} basically tells that the maximum of $M_t$ given by (\ref{pop-mean}) is attained at one of the change-points $1 \le \tau_1 < \cdots < \tau_q <T$. 

\setcounter{lemma}{2}
\begin{lemma}
\label{lem3}
 Let $1 \le \tau_1 < \cdots < \tau_q <T$ be $q \ge 1$ change-points such that $\mu_1=\cdots=\mu_{\tau_1}\ne \mu_{\tau_1+1}=\cdots=\mu_{\tau_q}\ne \mu_{\tau_q+1}=\cdots=\mu_T$. 
 Then, $M_t$ defined by (\ref{pop-mean}) attains its maximum at one of the change-points. 
\end{lemma}

Now let's prove Theorem 5. Recall that within the time interval $[1, T]$, there are $q$ change-points. First, we will show that the proposed binary segmentation algorithm detects the existence of change-points with probability one. To show this, according to Theorem 3, we only need to show that $\mbox{P}(\hat{\mathscr{M}} [1, T]>\mathscr{M}_{\alpha_n}[1, T])=1$ where $\mathscr{M}_{\alpha_n}$ is the upper $\alpha_n$ quantile of the Gumbel distribution. This can be shown because for any $1 \le t \le T-1$,
\begin{eqnarray}
\mbox{P}(\hat{\mathscr{M}} [1, T]>\mathscr{M}_{\alpha_n}[1, T]) &\ge& \mbox{P}(\frac{\hat{{M}_t}}{\sigma_{nt,0}}>\mathscr{M}_{\alpha_n}[1, T])\\
&=&1-\Phi(\frac{\sigma_{nt,0}}{\sigma_{nt}}\mathscr{M}_{\alpha_n}-\frac{M_t}{\sigma_{nt}}), \nonumber
\end{eqnarray}    
which converges to 1 because $\sigma_{nt,0} \le \sigma_{nt}$, $M_t/\sigma_{nt} \ge \mathscr{M}^* \to \infty$, and $\mathscr{M}_{\alpha_n}=o(\mathscr{M}^*)$.  

Once the existence of change-points is detected, the proposed binary segmentation algorithm will continue to identify change-points. Since $v_{\max}=o\{n \delta/ (T \sqrt{\log T}) \}$, one change-point $\tau_{(1)} \in \{\tau_1, \cdots, \tau_q \}$ can be identified correctly with probability 1 based on similar derivations given in the proof of Theorem 4, and the fact that $M_t$ achieves its maximum at one of change-points as shown in Lemma 3. 

Since each subsequence satisfies the condition that $\mathscr{M}_{\alpha_n}=o(\mathscr{M}^*)$, the detection continues. Suppose that there are less than $q$ change-points identified successfully, then there exists a segment $I_t$ contains a change-point. Since $\mathscr{M}_{\alpha_n}=o(\mathscr{M}^*)$ and $v_{\max}[I_t]=o\{n \delta[I_t]/ (T \sqrt{\log T}) \}$, the change-point will be detected and identified by the proposed binary segmentation method. Once all $q$ change-points have been identified consistently, each of all the subsequent segments has two end points chosen from $1, \tau_1, \cdots, \tau_q, T$. Then the proposed binary segmentation algorithm will not wrongly detect any change-point from any segment $I_t$ that contains no change-point, because according to Theorem 3, 
$
\mbox{P}(\hat{\mathscr{M}} [I_t]>\mathscr{M}_{\alpha_n}[1, T])=\alpha_n \to 0,
$         
which implies that no change-point will be identified further. This completes the proof of Theorem 5.

\end{document}